\documentclass[%
 reprint,
 superscriptaddress,
 amsmath,amssymb,
 aps,
 prb,
]{revtex4-2}

\setcitestyle{number}
\usepackage{graphicx}
\usepackage{amsmath,amsfonts,amssymb}
\usepackage[dvipsnames]{xcolor}
\definecolor{MyGreen}{RGB}{69,139,0}
\usepackage{dcolumn}
\usepackage{bm}
\usepackage{hyperref}
\hypersetup{
    colorlinks=true,
    linkcolor=blue,
    filecolor=magenta,      
    urlcolor=blue,
    citecolor=blue
    }
\usepackage{upgreek}
\usepackage{ragged2e}

\usepackage{blindtext}

\usepackage{etoolbox} 

\makeatletter
\renewcommand{\l@section}{\@dottedtocline{1}{1.5em}{2.3em}}
\renewcommand{\l@subsection}{\@dottedtocline{2}{3.8em}{3.2em}}
\renewcommand{\l@subsubsection}{\@dottedtocline{3}{7.0em}{4.1em}}
\makeatother

\let\oldtableofcontents\tableofcontents

\renewcommand{\tableofcontents}{%
  \begingroup 
    \hypersetup{hidelinks} 
    \oldtableofcontents
  \endgroup 
  \hypersetup{colorlinks=true, linkcolor=blue, citecolor=blue, urlcolor=blue} 
}

\makeatletter
\renewcommand{\p@subsection}{}   
\renewcommand{\p@subsubsection}{} 
\makeatother

\makeatletter
\renewcommand{\l@section}{\@dottedtocline{1}{1.5em}{1.5em}}
\renewcommand{\l@subsection}{\@dottedtocline{2}{3.8em}{2em}}
\renewcommand{\l@subsubsection}{\@dottedtocline{3}{7.0em}{3em}}
\makeatother

\begin{document}


\title{The Reemergence of Selenium Solar Cells}

\author{Rasmus S. Nielsen}
\email[]{Electronic mail: raniel@dtu.dk}
\affiliation{SurfCat, Department of Physics, Technical University of Denmark, 2800 Kongens Lyngby, Denmark}

\begin{abstract}

Selenium, the world's oldest photovoltaic material, has experienced a renaissance in research over the past decade, with certified solar cell efficiencies climbing from the historical record of 5\% to breaking the 10\% barrier. Its wide bandgap makes it a particularly interesting candidate for tandem solar cells and indoor photovoltaic applications, yet despite steadily improving the carrier collection, devices consistently suffer from a substantial open-circuit voltage deficit. This review provides a critical analysis of the material properties and optoelectronic quality of state-of-the-art selenium thin films. Published results from independent groups are digitized and directly compared, collectively painting a comprehensive picture of the carrier dynamics, supported and contextualized by drift-diffusion simulations. Strategies for synthesizing and processing selenium thin films are also examined in detail, highlighting not only best practices but also the underlying crystal growth kinetics that ultimately govern material quality. Finally, a series of open questions and challenges is presented, spanning from fundamental materials science and atomic-scale defect physics to device-level engineering, providing a roadmap to unlock the intrinsic photovoltaic potential of selenium and guide the future development of higher-efficiency selenium solar cells.

\end{abstract}

\maketitle

\tableofcontents

\begin{figure*}[t!]
    \centering
    \includegraphics[width=\textwidth,trim={0 0 0 0},clip]{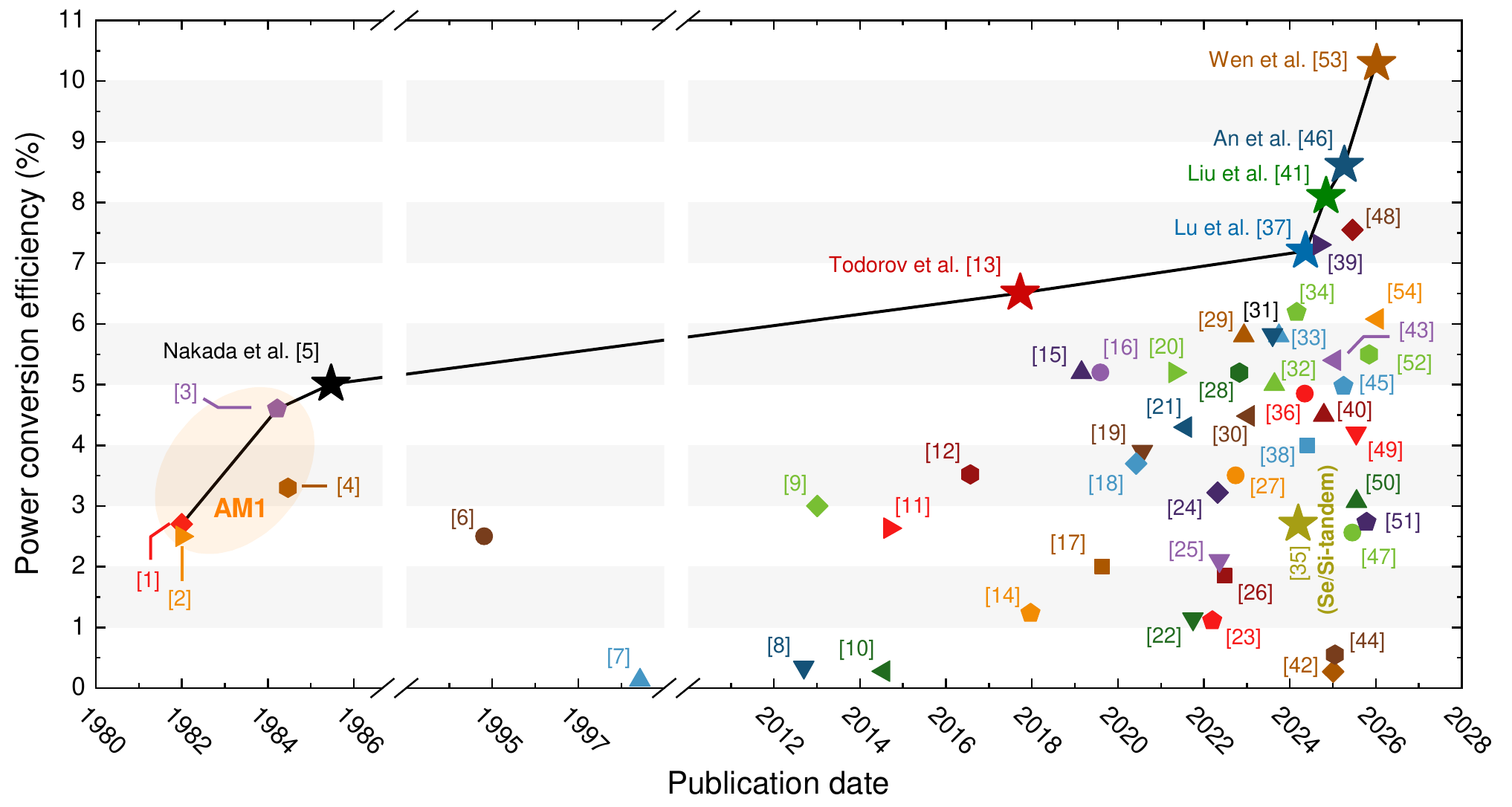}
    \caption{Reported power conversion efficiencies (PCE) of selenium solar cells as a function of publication year (last updated: 07-01-2026). Each data point is labeled with its corresponding reference \cite{ito1982a, kunioka1982a, ito1984a, nakada1984a, nakada1985a, champness1995a, tennakone1998a, qian2012a, nguyen2013a, panahi-kalamuei2014a, wang2014a, zhu2016a, todorov2017a, zhu2018a, hadar2019a, hadar2019b, wu2019a, liu2020a, liu2020b, youngman2021a, liu2021a, nielsen2021a, deshmukh2022a, fu2022a, thomas2022a, zheng2022a, lu2022a, nielsen2022a, yan2022a, kawagishi2023a, liu2023a, nielsen2023a, wei2023a, wang2024a, nielsen2024a, li2024a, lu2024a, nielsen2024b, lu2025a, kobayashi2024a, liu2025a, huang2025a, bao2025a, singh2025a, nielsen2025a, an2025a, wang2025a, shen2025a, ca2025a, kim2025a, alfieri2025a, segura-blanch2025a, wen2026a, duan2026a}, and record efficiencies are marked with star symbols. Devices tested under AM1 illumination in the early 1980s are shown within the orange shaded region, distinguishing them from later reports measured under standard AM1.5G conditions.}
    \label{fig:TimelinePCE}
\end{figure*}

\section{Introduction}

The elemental semiconductor selenium was the first material ever used to convert light into electricity, laying the foundation for a range of optoelectronic technologies long before the emergence of modern semiconductor materials. It played a central role in the discovery of photoconductivity and the photovoltaic effect, and was used as the absorber material in the very first solar cell. Back then, however, solar cells were mostly a lab curiosity, and by the time the space race created an urgent demand for practical solar power, silicon — the new wonder-material of the microelectronics age — had already established its dominance. With its more advanced, higher-quality manufacturing and a lower bandgap better suited for single-junction solar cells, silicon stole the spotlight from selenium, and interest in selenium photovoltaics declined sharply. This first wave of selenium solar cell development culminated in a record efficiency of 5\% reported by Nakada et al. in 1985 \cite{nakada1985a}.

Today, the efficiency of silicon solar cells has reached 27.9\% \cite{green2025a}, approaching the upper limit of single-junction devices \cite{richter2013a}. Therefore, to further advance the performance of photovoltaics (PV), research must focus on higher-efficiency concepts, such as tandem devices and indoor photovoltaics (IPV). These next-generation PV technologies rely on wide-bandgap photoabsorbers in the range of 1.6-2.0 eV, yet only a limited number of candidates have been proposed, none of which have simultaneously demonstrated high performance, low cost and long-term stability. The search for an ideal wide-bandgap absorber has revitalized research on selenium, particularly following the work by Todorov \textit{et al}. in 2017 \cite{todorov2017a}. Since then, rapid progress has been made, with reported device efficiencies rising from the historical record of 5\% to beyond 10\% in less than a decade (see Fig.~\ref{fig:TimelinePCE}).

Selenium thin film solar cells have been fabricated with a wide range of different device architectures using different techniques, but several key challenges remain. In particular, state-of-the-art devices consistently exhibit an open-circuit voltage deficit exceeding 550 mV, representing the primary limitation on PV performance. This raises important questions about the optoelectronic quality of selenium thin films across different research groups. To better understand the origins of this loss and enable the growth of higher-quality absorbers, it is essential to examine both the structural chemistry and the anisotropic material properties relevant for photovoltaics. This includes electronic and vibrational band structures, defect tolerance, absorption coefficient, radiative efficiency, carrier mobilities, and lifetimes, alongside device-level considerations such as band alignment, interfaces, and carrier-selective contact layers, using both experimental and simulation tools. To paint a more comprehensive picture of material properties and carrier dynamics in selenium solar cells, published results from independent groups are digitized and directly compared, with hypotheses and conclusions drawn from the collective research landscape.

As interest in selenium photovoltaics continues to grow, this review aims to critically assess the latest developments and highlight future directions. To address common inaccuracies in the literature, scientific insights are framed within a concise historical context, while considerations of scalability are discussed to evaluate the commercial and technological relevance of selenium solar cells. By examining these interconnected aspects, this review identifies key open scientific questions spanning from fundamental science at the atomic scale to engineering challenges in complete photovoltaic device. The goal is to provide a coherent framework that can guide future research and support the development of higher-performance selenium PV devices.

\section{The History of Selenium Photovoltaics}

The origins of photovoltaics date back to 1839, when Alexandre Edmond Becquerel discovered the photovoltaic effect in an electrochemical cell. He observed that illuminating a silver chloride electrode in an acidic solution generates a measurable photovoltage, marking the first demonstration of light-to-electricity conversion \cite{becquerel1839a}. The next major milestone came with the discovery of photoconductivity in solids by Willoughby Smith and Joseph May, who showed that the electrical resistance of selenium changes upon illumination \cite{Smith1873}. Shortly thereafter, in 1876, William Grylls Adams and Richard Evans Day illuminated a rod of sintered selenium contacted with platinum electrodes and generated electrical power without any applied bias \cite{adams1877a}. Although highly inefficient and impractical, this experiment constitutes the first demonstration of a solid-state photovoltaic device. Some reports suggest that these early studies on the photophysics of selenium inspired Einstein’s Nobel Prize-winning work, but the photovoltaic effect should not be confused with the photoelectric effect, which was first observed by Heinrich Hertz in 1887 \cite{hertz1887a}.

In 1883, Charles Fritts fabricated another type of selenium solar cell by coating a metal substrate with thin layers of selenium and gold \cite{fritts1883a}. This device had a surface area of approximately 30 cm$^\text{2}$ and a power conversion efficiency approaching 1\%, and it is commonly recognized as the first solar cell ever made. Although technologically more advanced and practical -- and indeed the first \textit{thin film} solar cell -- the demonstration of the first solid-state photovoltaic device should arguably be credited to Adams and Day. Nevertheless, Werner von Siemens recognized the significance of Fritts’ work, declaring to the Royal Academy of Prussia that the direct conversion of light into electricity had been demonstrated \cite{mertens2018a}.

Photovoltaics was for the most part just a scientific curiosity until the 1950s, when Bell Laboratories began exploring the potential of solar energy for telecommunications. Building on William B. Shockley’s explanation of the working principles of the pn-junction in 1949 \cite{shockley1949a}, Chapin, Fuller, and Pearson developed the first silicon solar cell in 1954, achieving very promising efficiencies of up to 6\% \cite{chapin1954a}, which rose to nearly 10\% by 1958. However, the high cost of the technology meant that over the following two decades, advances in silicon and III-V photovoltaics were largely driven by the space race \cite{mertens2018a}.

The oil crisis in 1973 shifted our attention from fossil fuels to alternative energy sources, and terrestrial solar power became the center of interest. Silicon, with its lower bandgap and more advanced processing, stole the spotlight from selenium, and research on selenium-based PV culminated in a final efficiency milestone of 5\% reported by Nakada and co-workers in 1985 \cite{nakada1985a}. Progress in selenium solar cells stalled for decades until 2017, when Todorov and co-workers achieved a new record efficiency of 6.5\% and highlighted its promise as both a wide-bandgap absorber for multi-junction solar cells and an ideal candidate for indoor photovoltaics (IPV) \cite{bishop2017a}.

\section{Abundance, Scalability and Toxicity}

Selenium is often described as an Earth-abundant and non-toxic absorber material, a claim that is commonly used to motivate the exploration of its potential in photovoltaics. However, evaluating the true scalability of selenium-based PV requires a broader perspective that considers not only crustal abundance and toxicity, but also production capacity, supply elasticity, extraction efficiency, and recyclability.

The average crustal abundance of selenium is about $\sim$0.05~ppm \cite{yang2021a}, roughly on par with indium, which is considered a critical material and a potential bottleneck for scaling other thin film PV technologies such as CIGS \cite{andersson2000a}. However, a more meaningful measure of scalability is the availability and elasticity of production capacity. Selenium is produced mainly as a by-product of the electrolytic refining of copper, where it is recovered from anode slimes. As a result, the supply of selenium is closely tied to global copper production and is inherently inelastic. In 2025, the global production of selenium was estimated at around 3800~t/yr \cite{USGS_MCS_2026}. The extraction efficiencies from anode slimes are close to 60\%, but it is considered highly unlikely to further improve \cite{green2006a}. Other potential sources of selenium include lead, nickel, and zinc ores, or even coal, but recovering selenium from these sources is either limited in scale or not economically feasible. This leaves little room for expanding production significantly, even if prices were to rise \cite{vesborg2012a}.

A rough estimate of the upper limit for scaling selenium-based PV can be made with a simple back-of-the-envelope calculation. Trigonal selenium has a density of 4.81~g/cm$^\text{3}$, corresponding to an areal mass of 4.81~g/m$^\text{2}$ for a 1~$\upmu$m thick absorber. Assuming a device efficiency of 25\%, full allocation of the global production of selenium ($\sim$3800~t/yr), and no material losses during production, the current supply could support an annual installed PV capacity of roughly 200~GWp/yr. That may sound impressive, but it is clearly an extreme upper bound: selenium is also needed in agriculture, glass production, electronics, chemicals, and pigments \cite{USGS_MCS_2026}, and not only are current record efficiencies below 11\% \cite{wen2026a}, but only a fraction of the source material actually makes it into the finished device. Realistically, the inelastic supply of selenium would likely support only moderate-scale deployment of selenium-based PV.

\begin{figure*}[t!]
    \centering
    \includegraphics[width=\textwidth,trim={0 0 0 0},clip]{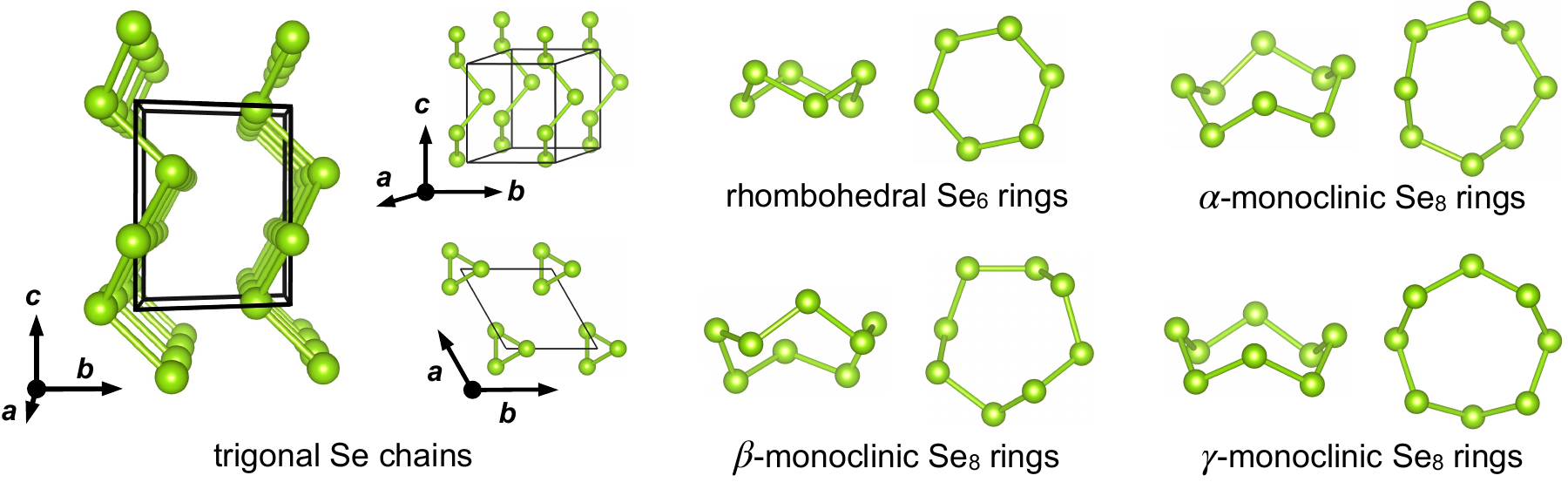}
    \caption{Crystal structures of elemental selenium in its trigonal, rhombohedral, and monoclinic ($\alpha$, $\beta$, and $\gamma$) phases. The trigonal structure is shown along different crystallographic directions to highlight its anisotropic bonding nature. All crystal structures were visualized in VESTA using crystallographic information files (CIFs) sourced from the ICSD database.}
    \label{fig:CrystalStructures}
\end{figure*}

Recyclability is another important consideration for the life cycle and sustainability assessment of emerging technologies \cite{woods-robinson2025a}. In 2024, Wang et al. demonstrated a straightforward method to recover selenium from end-of-life selenium solar cells by exploiting its high vapor pressure relative to the other device components \cite{wang2024a}. Their process achieved recycling yields of about 98\%, enabling the refabrication of photovoltaic devices with performance comparable to those made from commercially sourced materials. This highlights the potential for closed-loop material use and suggests that recycling could help offset supply constraints.

In my opinion, claims of non-toxicity are sometimes thrown around a little too freely when discussing emerging materials. Selenium is an essential trace element for humans, with an upper tolerable intake of roughly 400~$\upmu$g/day \cite{macfarquhar2010a}, but its toxicity depends strongly on the chemical form and bioavailability \cite{wilber1980a}. Elemental selenium and most solid-state selenides are poorly absorbed in the human body, whereas soluble selenium species, organoselenides and hydrogen selenide (H$_\text{2}$Se) gas are highly toxic and corrosive, meaning that certain processing steps could pose environmental and/or health concerns \cite{olson1987a}. However, in solid-state photovoltaic devices, particularly when encapsulated, selenium generally presents low environmental and health risks. Nonetheless, it is more accurate to describe selenium as a low-toxicity, rather than a truly non-toxic, absorber material.

Although the inelastic production capacity of selenium limits the prospects for terawatt-scale PV, the commercial success of CdTe photovoltaics and the ongoing upscaling of lead-halide perovskites offer a useful perspective. In 2025, First Solar sold a volume of 17.5~GWp of CdTe modules \cite{firstsolar2025annual}, despite tellurium being scarcer than selenium and CdTe layers being much thicker than typical selenium absorbers. Furthermore, the toxicity and environmental concerns associated with Pb, Cd, and Te are considerably greater than for selenium, demonstrating that Earth-abundance and toxicity constraints alone do not preclude technological relevance. They remain important factors when assessing market potential and opportunities, and should be considered accordingly.

\section{Structural Chemistry}\label{sec:StructuralChemistry}

Elemental selenium exists in several low-energy allotropes in the solid state, reflecting its ability to concatenate into ring- and chain-like structures. These allotropes are generally classified as either crystalline or amorphous, with the crystalline phases shown in Fig.~\ref{fig:CrystalStructures}. The chain-like trigonal phase (space groups P3$_\text{1}$21 and P3$_\text{2}$21, corresponding to right- and left-handed screw directions, respectively \cite{ramirez-montes2024a}) is thermodynamically the most stable and, due to its favorable material properties, is the preferred structure for photovoltaic applications. The metastable ring-like crystalline phases are also important to recognize, as they dominate the vapor phase used in the most common deposition routes for synthesizing selenium thin films. Finally, although amorphous selenium is not discussed in detail here, it is worth noting that its atomic structure remains controversial and continues to be debated in the literature \cite{marple2017a, lu2024b}.

The molecular chemistry of elemental selenium in the gas phase closely mirrors that of sulfur. While selenium forms covalently bonded chains in the trigonal phase, these extended structures break down when thermal energy is supplied for sublimation or evaporation. Upon heating, the chains fragment and preferentially cyclize into molecular rings, with ($\alpha$, $\beta$, and $\gamma$)-monoclinic Se$_\text{8}$ being the most thermodynamically stable species. This preference arises from a balance between bond energetics and ring strain, favoring Se$_\text{8}$ over smaller or larger cyclic oligomers at moderate temperatures \cite{gleiter2019a}. As a result, selenium evaporates predominantly as Se$_\text{8}$ under typical sublimation and evaporation conditions, with only minor contributions from other molecular species. Smaller fragments such as Se$_\text{2}$ or atomic selenium appear only at significantly higher temperatures, where entropic effects drive further dissociation. Consequently, vapor-phase deposition routes proceed through these metastable, closed-shell molecular precursors that differ markedly from the chain-like trigonal selenium and are not particularly reactive upon reaching the substrate. The reactivity of impinging selenium species can be increased by raising the substrate temperature or by cracking the molecular beam using heating or a plasma, which promotes Se-Se bond cleavage and allows for more reactive selenium species to form covalent bonds with the substrate. These approaches are particularly interesting, as they enable preferential growth of trigonal selenium thin films, which will be discussed further in Section~\ref{sec:FilmTexture}. While increasing the temperature of the crucible would also generate more reactive species, this simultaneously raises the evaporation rate, which is generally undesirable in the context of epitaxial growth of thin films.

\begin{figure*}[t!]
    \centering
    \includegraphics[width=\textwidth,trim={0 0 0 0},clip]{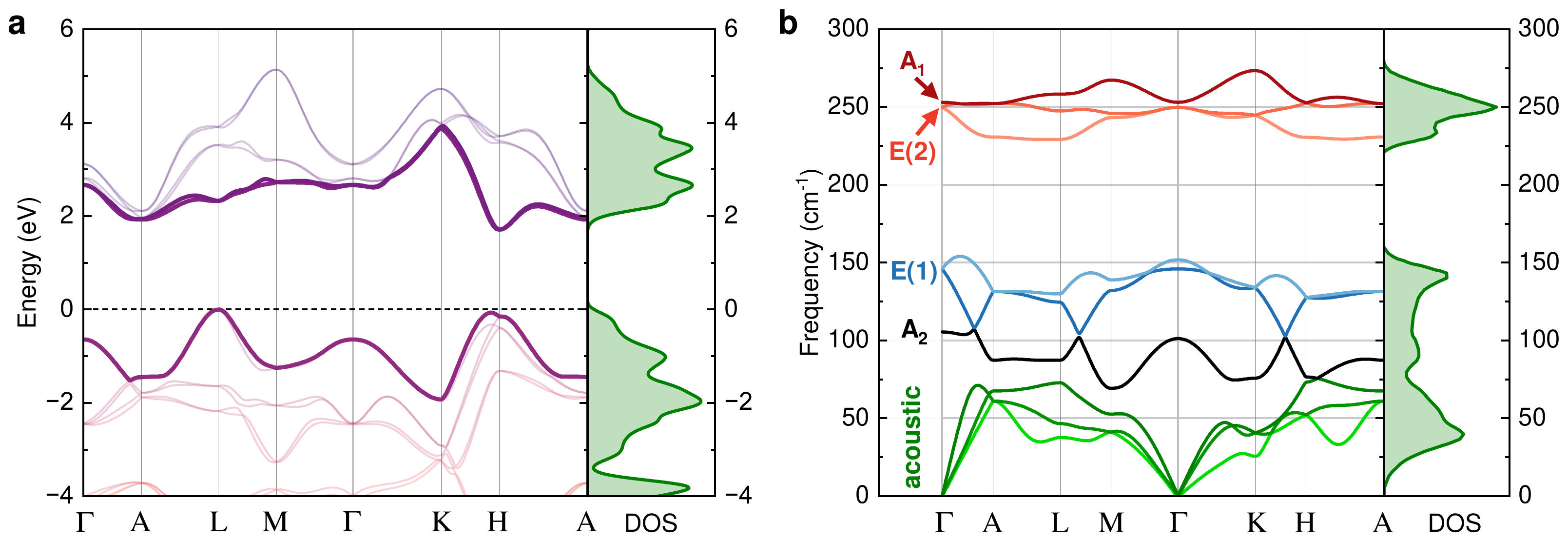}
    \caption{Bulk (a) electronic and (b) phonon band structures, alongside the density of states (DOS), of trigonal selenium, calculated using hybrid density functional theory (HSE06) including spin-orbit coupling and van der Waals dispersion corrections, as reported by Kavanagh \textit{et al.} \cite{kavanagh2025a, nielsen2025b}.}
    \label{fig:BandStructures}
\end{figure*}

Trigonal selenium is often referred to as a quasi-1D material due to its anisotropic crystal structure, in which the covalent Se-Se bonds that make up the helical chains are aligned along one crystallographic direction, and adjacent chains are held together only by much weaker van der Waals forces. This structural anisotropy gives rise to highly anisotropic material properties, making control over crystallographic orientation key to maximizing device performance. In photovoltaic devices, it is more specifically desirable to align the \textit{c}-axis (chain direction) with the direction of charge transport, which motivates the aforementioned cracking of molecular selenium species to form anchoring covalent bonds with the substrate. Selenium shares this quasi-1D covalent bonding character with a family of emerging low-dimensional photovoltaic materials, including Sb$_\text{2}$(S,Se)$_\text{3}$ \cite{shah2021a, chen2022a}, Sn(S,Se)/Ge(S,Se) \cite{mitzi2022a, cho2020a, sinsermsuksakul2014a, liu2021b, liu2020c}, and (Sb,Bi)(S,Se)(Br,I) \cite{zhang2025a, cano2025a, nielsen2025c}. A key advantage of this low-dimensional nature is that grain boundaries perpendicular to the chains are expected to be benign due to the low electronic band dispersion in this direction, meaning that high-performance devices do not necessarily require large \textit{single-crystal} thin films. Such benign grain boundaries have also been observed experimentally in both Sb$_\text{2}$Se$_\text{3}$ \cite{zhou2015a} and selenium \cite{wu2025a} using Kelvin probe force microscopy.

Tellurium is another important member of this materials family, as it crystallizes in the same structure as selenium and is therefore a so-called \textit{isostructural} material. Like sulfur, the gas phase chemistry of tellurium also closely resembles that of selenium, and tellurium has played several important roles in the development of selenium solar cells. Most notably, an ultra-thin layer of tellurium is more often than not used to promote the surface wetting of selenium, which is otherwise challenging due to its low surface energy in combination with its high vapor pressure. This will be discussed further in Section \ref{sec:SurfaceWetting}. In addition, the isostructural relationship between selenium and tellurium enables the formation of Se-Te solid solutions. This expansion of the chemical space introduces additional degrees of freedom for tuning optoelectronic material properties. For example, the optical bandgap can be tuned from 1.85~eV to 0.35~eV, as first demonstrated in the context of photovoltaic applications by Hadar \textit{et al.} \cite{hadar2019a}.

Trigonal selenium is intrinsically chiral due to its helical chain structure, which gives rise to left- and right-handed \textit{enantiomorphs} \cite{ramirez-montes2024a}. While there is no clear reason to expect chirality itself to directly affect the bulk material properties relevant to photovoltaics, it may still influence nucleation and crystal growth modes, particularly on anisotropic or epitaxial substrates. At present, it remains unclear from the literature whether individual grains consist exclusively of chains with one handedness or can contain a mixture of both. If chains of opposite handedness can coexist locally, boundaries between them may form enantiomorphic domain walls and, as with conventional grain boundaries in selenium, no covalent bonds would be broken. Such extended defects would therefore most likely also be electrically inactive, but they could still affect the local phonon dispersion and electron-phonon interactions, with potential consequences for both exciton dynamics and charge transport. At present, however, the formation and control of chirality during the growth of selenium thin films remain poorly understood.

\subsection{Band Structures}\label{sec:BandStructures}

The bulk electronic and phonon band structures of trigonal selenium are shown in Fig.~\ref{fig:BandStructures}a and b, respectively. Both were calculated using hybrid density functional theory (HSE06) including spin-orbit coupling and van der Waals dispersion corrections, as reported by Kavanagh \textit{et al.} \cite{kavanagh2025a}. The electronic band structure shows that the fundamental bandgap is in fact indirect, but with only a very small energy difference of $\Delta E_\text{g,direct/indirect}\simeq$~0.1~eV between the lowest indirect and direct transitions. This difference is so small that trigonal selenium is sometimes described as a quasi- or pseudo-direct bandgap material. Because indirect transitions are generally weak, the lowest direct transition lies very close in energy to the lowest indirect transition, and selenium absorber layers are typically thin, it is experimentally difficult to distinguish indirect absorption from the tail of the direct transition \cite{nielsen2022a}. Although the indirect edge could in principle reduce radiative recombination, this energy difference of $\Delta E_\text{g,direct/indirect}\simeq$~0.1~eV is expected to translate almost directly into an open-circuit voltage loss when non-radiative recombination dominates \cite{kirchartz2017a}. Another important feature is the pronounced non-parabolicity of the valence-band edge, as the effective density of states (DOS) and carrier effective masses, both of which are important for photovoltaic materials, are often estimated using the parabolic band approximation. These properties should therefore instead be assessed through Brillouin-zone integration, which will be discussed further in Section \ref{sec:DOSandMasses}.

The phonon dispersion curves shown in Fig.~\ref{fig:BandStructures}b consist of six phonon modes with the irreducible representation $\Gamma_\text{vib} = \mathrm{A}_\text{1} + 2\mathrm{A}_\text{2} + 3\mathrm{E}$. Of these, $\mathrm{A}_\text{2}+\mathrm{E}$ correspond to the acoustic modes, $\mathrm{A}_2+2\mathrm{E}$ are infrared-active, and $\mathrm{A}_\text{1}+\text{2}\mathrm{E}$ are Raman-active. Because the $\mathrm{A}_\text{1}$ and $\mathrm{E}(\text{2})$ modes are nearly degenerate in frequency at the $\Gamma$-point, they can be difficult to distinguish experimentally, although doing so is important for the correct interpretation of vibrational spectroscopic measurements \cite{nielsen2025b}. Another notable feature of the phonon band structure is the gap separating the low- and high-frequency optical branches. This gap restricts the available phonon scattering channels and may therefore suppress certain anharmonic decay pathways. In principle, this could be advantageous, as phonon-mediated carrier thermalization may be slower and radiative efficiency correspondingly higher. However, the optical phonon density of states is also concentrated within a relatively narrow range of high-energy bond-stretching modes, meaning that photo-excited carriers interact primarily with these vibrations. This may strengthen electron-phonon coupling and promote defect trapping and non-radiative recombination.

\subsection{Intrinsic Point Defect Tolerance}

\begin{figure}[ht!]
    \centering
    \includegraphics[width=\columnwidth,trim={0 0 0 0},clip]{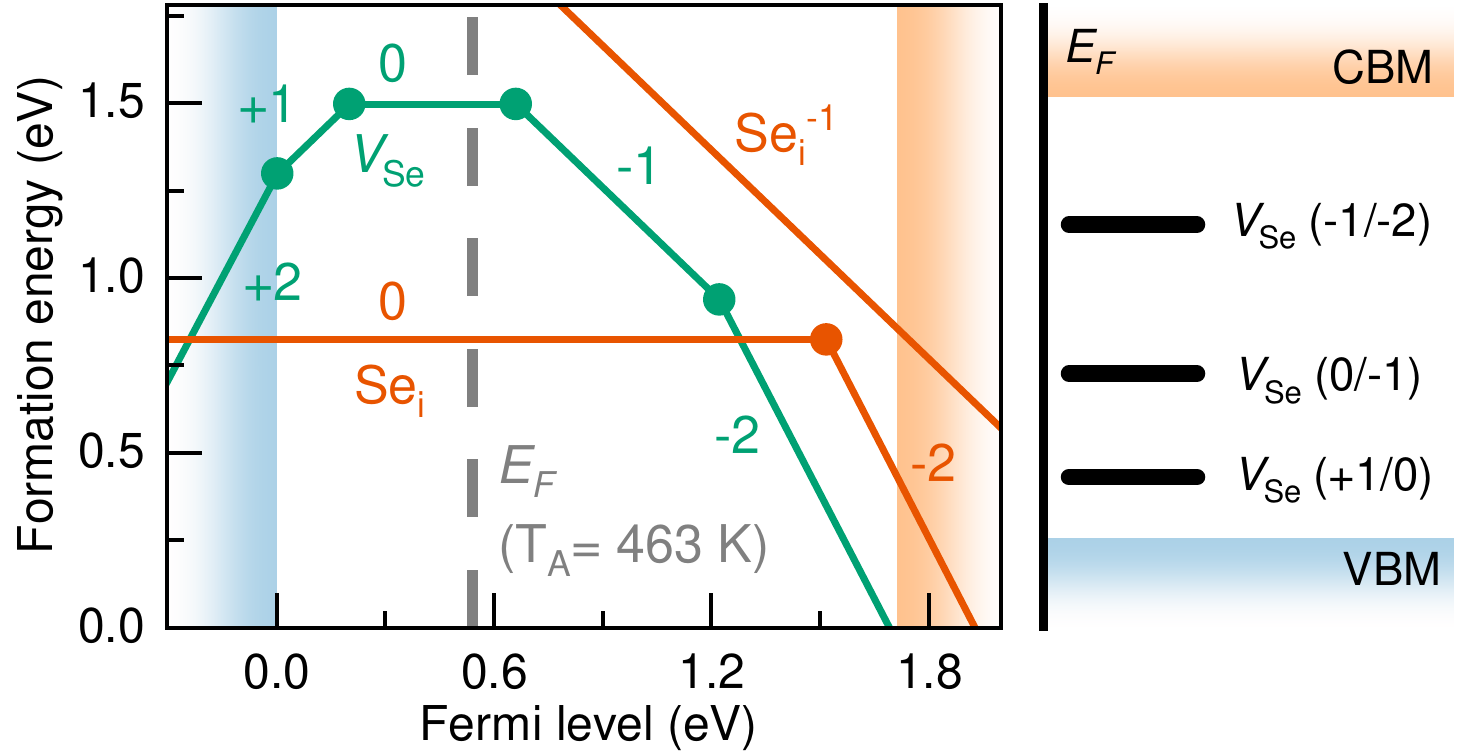}
    \caption{Formation energies of intrinsic point defects in trigonal selenium as a function of Fermi level position within the bandgap (left), alongside a schematic band diagram showing the charge transition levels of selenium vacancies ($V_\text{Se}$) (right). As reported by Kavanagh \textit{et al.} \cite{kavanagh2025a}\vspace{-0.5cm}}
    \label{fig:IntrinsicDefects}
\end{figure}

One advantage of a single-element system is that selenium has only two possible native point defects, namely vacancies ($V_\text{Se}$) and interstitials ($\text{Se}_\text{i}$). As the equilibrium population of native defects often sets an intrinsic limit on the achievable efficiency of emerging photovoltaic materials, it is important to assess whether these defects introduce deep charge transition levels in the bandgap, and to quantify their carrier-capture coefficients and associated recombination rates. Since the intrinsic defect chemistry is difficult to probe experimentally, first-principles methods based on density functional theory (DFT) have been widely used, and several studies have already examined the defect chemistry of trigonal selenium, including both intrinsic and extrinsic point defects \cite{kavanagh2025a, moustafa2024a, li2024b}. The conclusions of these studies are consistent. Of the two intrinsic point defects, selenium self-interstitials have the lowest formation energy, but they are inactive with respect to both doping and electron-hole recombination and are therefore considered benign. Selenium vacancies, by contrast, introduce several charge transition levels deep in the bandgap and would therefore be expected to contribute to both doping and recombination. However, their medium-to-high formation energies suggest relatively low equilibrium concentrations, and their calculated capture cross-sections are also low, implying low non-radiative recombination rates through this type of defect. The formation energies of the two intrinsic defects are shown as a function of the Fermi level in Fig.~\ref{fig:IntrinsicDefects}, alongside an energy band diagram showing the charge transition levels of the amphoteric selenium vacancies. As both intrinsic point defects are considered benign for recombination and electrically inactive in the bulk, trigonal selenium can be regarded as intrinsically tolerant to native point defects.

\begin{figure*}[t!]
    \centering
    \includegraphics[width=\textwidth,trim={0 0 0 0},clip]{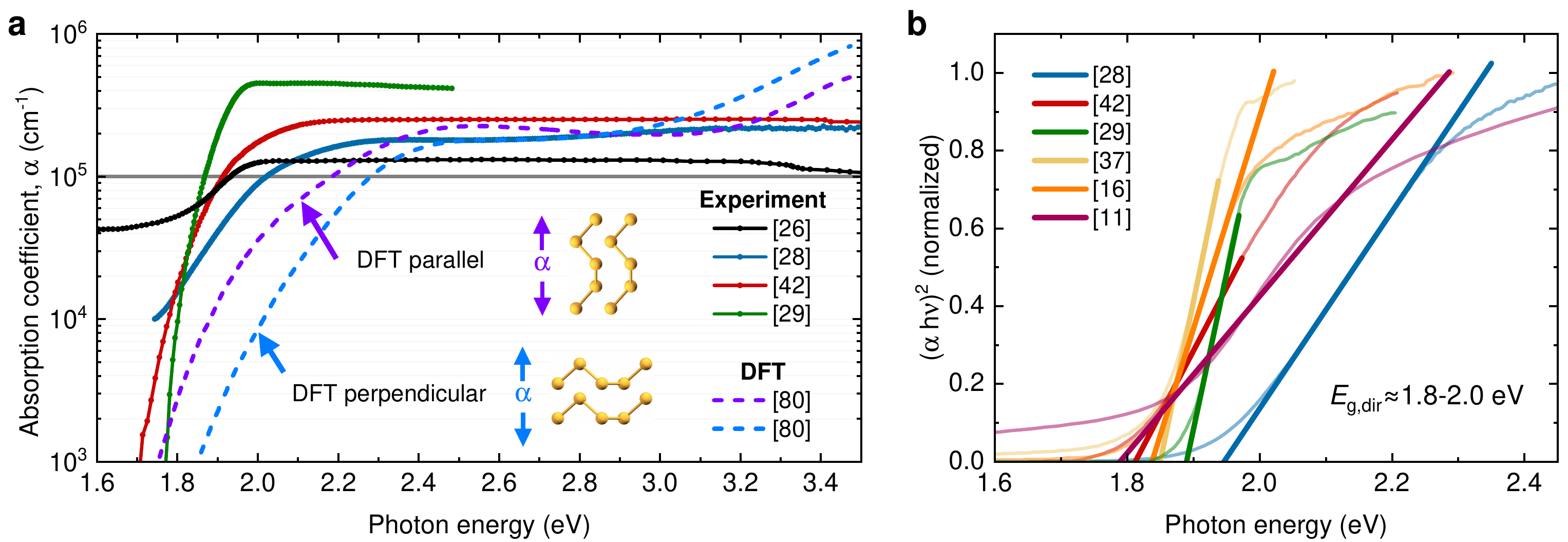}
    \caption{(a) Absorption coefficients of trigonal selenium thin films compiled from UV–vis spectroscopy measurements reported in Refs.~\cite{nielsen2022a, huang2025a, yan2022a, zheng2022a}, along with directionally dependent absorption coefficients calculated using DFT \cite{kavanagh2025a}, shown as dashed lines. (b) Tauc plots digitized, normalized, and shown with linear extrapolations used to extract optical bandgaps, compiled from Refs.~\cite{nielsen2022a, huang2025a, lu2024a, hadar2019b, yan2022a, wang2014a}.}
    \label{fig:AbsFigure}
\end{figure*}

\section{Material Properties}

The relevance of a photovoltaic absorber material is often judged primarily by its bandgap energy, which determines its suitability for single- or multi-junction solar cells. The corresponding upper efficiency limit is commonly estimated using the Shockley-Queisser theory \cite{shockley1961a}, but this framework provides only limited insights into the material properties that ultimately govern its true efficiency potential. More advanced models account for the effects of non-radiative recombination and finite diffusion lengths by including the absorption coefficient, carrier mobility and carrier lifetime \cite{kirchartz2018a}. In recent works by A. Crovetto \cite{crovetto2024a, crovetto2024b}, a phenomenological figure-of-merit (FOM) was proposed that also captures imperfect carrier collection by quantifying the influence of doping density and the dielectric constant on the PV performance potential. This dimensionless FOM is expressed as a function of eight bulk material properties,

\begin{equation}\label{eq:FOM}
    \Gamma_\mathrm{PV}= f\left(\alpha, \sigma, \tau, \mu, n, \epsilon, m, E_\mathrm{g}\right)
\end{equation}

\noindent where $\alpha$ and $\sigma$ describe the spectral average and dispersion of the absorption coefficient, $\tau$ is the non-radiative recombination lifetime, $\mu$ is the carrier mobility, $n$ is the doping density, $\epsilon$ is the static dielectric constant, $m$ is the density-of-states effective mass, and $E_\mathrm{g}$ is the bandgap.

This section critically assesses reported values of these bulk material properties for trigonal selenium, with particular emphasis on thin film measurements, as properties derived from single crystals and powders do not necessarily translate directly to device-relevant films. First-principles calculations have also been included as these properties can be difficult to assess reliably using only experimental characterization tools.

\subsection{Absorption Coefficient}

The absorption coefficient of trigonal selenium thin films is consistently reported to exceed $\alpha \geq 10^{5}\,\mathrm{cm}^{-1}$ in the visible spectral region across multiple independent studies. Such strong absorption is highly advantageous, as it enables efficient photon harvesting in ultra-thin absorbers. Notably, Todorov \textit{et al.} demonstrated that a film thickness of only $\sim$100 nm is sufficient to achieve an external quantum efficiency of approximately 80\% without the use of anti-reflective coating, which is an order of magnitude thinner than many established thin film absorber technologies, including CdTe, CIGS and halide perovskites. This also reduces material usage per unit area, which is beneficial from a scalability perspective.

\begin{figure*}[t!]
    \centering
    \includegraphics[width=\textwidth,trim={0 0 0 0},clip]{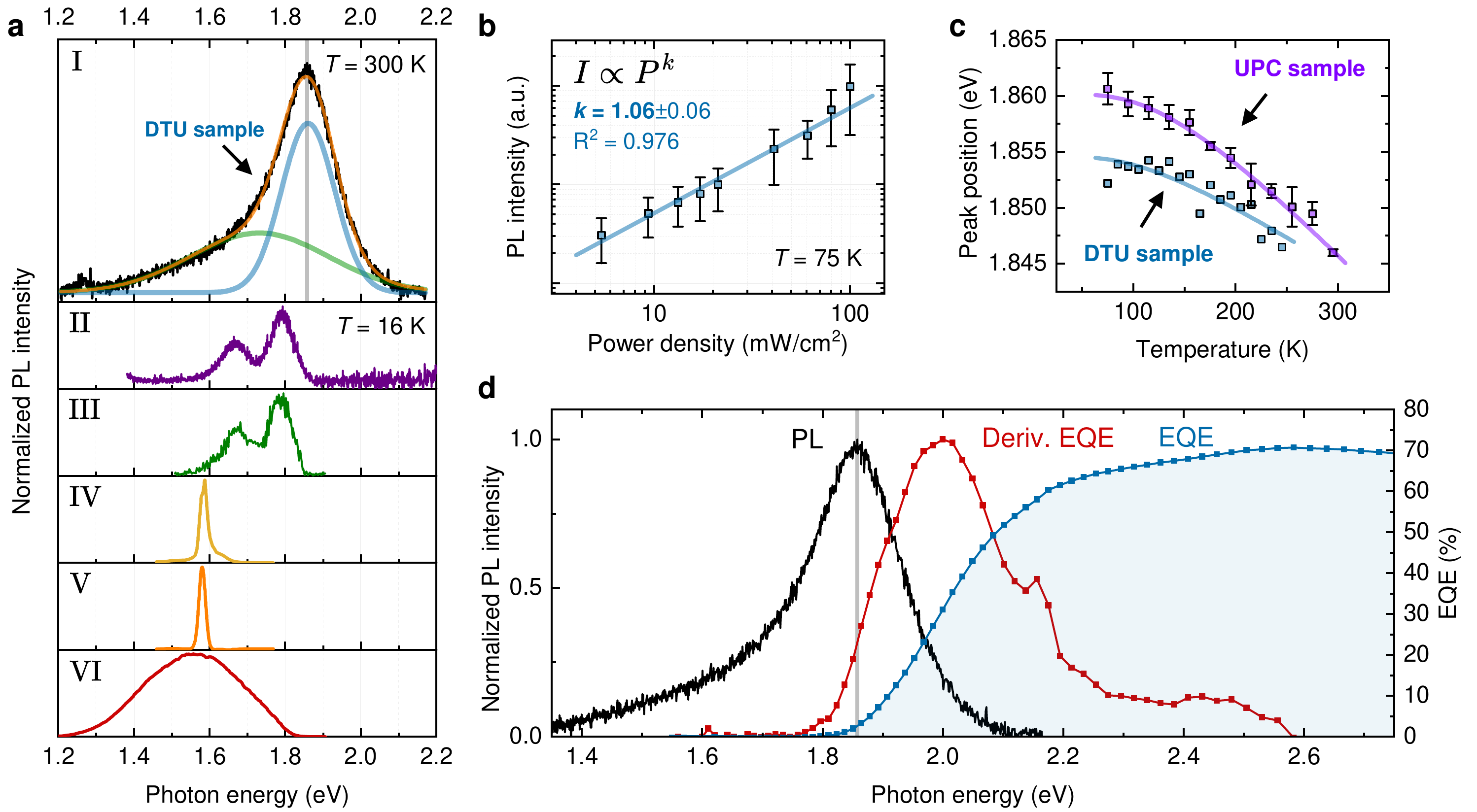}\newpage
    \caption{Photoluminescence characterization of selenium thin films. (a) Stack plot of digitized steady-state PL spectra from Refs.~\cite{nielsen2025b, nielsen2022a, liu2023a, li2024a, wu2019a, deshmukh2022a}, labeled I–VI. (b) Integrated PL peak intensity as a function of excitation power density at $T=75$~K, fitted to a power-law with exponent $k=1.06\pm0.06$, consistent with a band-to-band transition. (c) Temperature-dependent PL peak positions for samples grown at DTU and UPC, fitted with Bose-Einstein relations to quantify the electron-phonon coupling strength, showing weaker coupling for the DTU sample. (d) PL spectrum overlaid with an EQE spectrum and its derivative from a selenium solar cell from the same lab, highlighting the shift between optical and photovoltaic bandgaps.}
    \label{fig:PLfigure}
\end{figure*}

To enable a better and more direct comparison, reported absorption spectra from Refs.~\cite{zheng2022a, nielsen2022a, huang2025a, yan2022a} have been digitized and compiled in Fig.~\ref{fig:AbsFigure}(a). While all datasets confirm strong absorption, they also reveal substantial variation in the apparent onset energy and spectral dispersion near the absorption edge. Although all studies use UV–vis transmission spectroscopy, only one report explicitly accounts for diffuse reflection and transmission using an integrating sphere. This was shown to significantly affect both the steepness and position of the absorption onset in rough selenium thin films \cite{nielsen2022a}. First-principles calculations from Kavanagh \textit{et al.} are also included in Fig.~\ref{fig:AbsFigure}(a), predicting a broader absorption onset, with the onset extending to lower photon energies for polarization parallel to the chains \cite{kavanagh2025a}. Although optical anisotropy is expected to average out in randomly oriented polycrystalline films, all experimental datasets exhibit sharper absorption onsets than predicted by DFT, which is encouraging for photovoltaic applications.

As discussed in Section \ref{sec:BandStructures}, the fundamental bandgap of trigonal selenium is indirect. However, as the indirect transition is very close in energy to the lowest direct transition, and is expected to be weak, the optical direct bandgap is often determined from the absorption onset extrapolated from a Tauc plot. To better compare the results across different studies, the Tauc plots and the corresponding linear extrapolations from Refs.~\cite{nielsen2022a, huang2025a, lu2024a, hadar2019b, yan2022a, wang2014a} have been digitized, normalized and compiled in Fig.~\ref{fig:AbsFigure}(b), yielding bandgap values in the range of 1.8-2.0 eV. Although variations due to growth conditions \cite{bhatnagar1985a}, tellurium incorporation \cite{hadar2019a}, and structural disorder \cite{nielsen2025b} have been reported, the spread in values suggests that the optical bandgap of trigonal selenium thin films cannot be determined with high accuracy using conventional transmission measurements alone.

The dielectric function of trigonal selenium provides a more complete insight into its optical response, but most studies focus on single crystals rather than thin films. In 1967, Tutihasi and Chen reported the real and imaginary parts of the dielectric function parallel and perpendicular to the chains from reflectivity measurements on single crystals, revealing a pronounced excitonic absorption feature at 1.95 eV for light polarized perpendicular to the chains \cite{tutihasi1967a}. Interestingly, this is in close proximity to both the optical bandgap extracted from transmission measurements and the photovoltaic bandgap inferred from quantum efficiency spectra. However, the key dielectric property entering the FOM in Eq.~\ref{eq:FOM} is the static dielectric constant, previously estimated to be $\epsilon_r \approx 7.8$ from oscillator fits to infrared spectra of single crystals at room temperature \cite{madelung2004a}. To the best of my knowledge, the only thin film study of the dielectric function, based on spectroscopic ellipsometry, was unable to reliably extract the dielectric constants due to surface roughness and the lack of preferential grain orientation required to account for optical anisotropy \cite{nielsen2022a}. More fundamentally, optical techniques such as ellipsometry and reflectivity measurements probe the high-frequency dielectric response and are therefore not well suited for determining the true static dielectric constant. Capacitance-based measurements would instead provide access to the low-frequency response required to accurately determine the static dielectric constant, but such studies have not yet been reported for selenium thin films.

\subsection{Bandgap and Photoluminescence}

Given the difficulty in accurately determining the optical bandgap of trigonal selenium from absorption measurements, photoluminescence (PL) spectroscopy provides an alternative approach. PL is an incredibly powerful technique for assessing the optoelectronic quality of semiconductors, as it provides information on radiative recombination pathways. However, only a handful of studies present PL data for selenium thin films, and the reported spectra vary significantly. To compare these results, Fig.~\ref{fig:PLfigure}(a) shows a stack plot of digitized steady-state PL spectra compiled from Refs.~\cite{nielsen2025b, nielsen2022a, liu2023a, li2024a, wu2019a, deshmukh2022a}. As the highest-energy emission peak provides a lower bound for the optical bandgap, spectrum I from Ref.~\cite{nielsen2025b} indicates $E_\mathrm{g}\approx 1.85$~eV, in good agreement with the values extracted from the Tauc plots in Fig.~\ref{fig:AbsFigure}(b). The other PL spectra exhibit additional features at lower photon energies. Spectra II and III show two sub-bandgap peaks that were observed only at cryogenic temperatures in Ref.~\cite{nielsen2022a}, while spectra IV and V exhibit a single emission peak that is significantly redshifted and unusually sharp for band-to-band transitions at room temperature. Such features may instead originate from excitonic transitions, characterized by relatively narrow linewidths, or from more localized luminescent defect states. It is also worth noting that spectra IV and V were measured on selenium thin films grown within a mesoporous oxide film, suggesting a possible influence of the scaffold structure.

Early PL studies of trigonal selenium from the 1960s and 1970s focused exclusively on bulk single crystals grown via vapor phase or melt growth and were all measured at cryogenic temperatures. Queisser \textit{et al.} reported polarization-dependent PL from selenium crystals submerged in liquid hydrogen, with sharp spectral lines attributed to bound excitons \cite{queisser1967a}. Zetsche \textit{et al.} observed pronounced phonon satellites in the temperature range 2–50 K \cite{zetsche1969a}, while Küchler \textit{et al.} performed electro-PL measurements under applied electric fields to probe impurity-related transitions \cite{kuehler1973a}. Later, Moreth observed sharp emission lines from indirect free-exciton decay at 1.4 K, providing clear evidence of the quasi-indirect nature of the bandgap \cite{moreth1979a}. These findings were further supported in 1985 by time-resolved PL (TRPL) and photo-induced absorption measurements, which successfully explained the observed non-radiative decay channels using a simple linear electron–phonon coupling model \cite{chen1985a}.

In contrast to the free-standing single crystals, selenium thin films are typically grown directly on substrates and are therefore more susceptible to strain, randomly oriented grains, and structural disorder, all of which can strongly affect the PL response. The large variability in reported PL spectra highlights the need for systematic studies under well-defined and carefully documented conditions, necessary for correctly assigning emission bands to band-to-band, excitonic, or defect-mediated transitions. In particular, the high vapor pressure of selenium presents a major practical challenge, as thin films are prone to sublimation and degradation during measurement, even when using low excitation power densities. This, combined with a low PL quantum yield, may partly explain the limited use of PL as a diagnostic tool for selenium thin films.

In 2026, we introduced a closed-space encapsulation strategy that enables both power- and temperature-dependent PL measurements while effectively suppressing sublimation and degradation \cite{nielsen2025b}. The integrated peak intensity from power-dependent PL at $T=75$~K, shown in Fig.~\ref{fig:PLfigure}(b), follows a simple power-law relation with an exponent of $k\approx 1$, consistent with a band-to-band transition. Temperature-dependent PL peak positions from samples grown with nominally similar methods in two independent labs are shown in Fig.~\ref{fig:PLfigure}(c). The sample from DTU exhibits a markedly weaker temperature dependence than the sample from UPC, indicating weaker short-range electron-phonon coupling, reduced structural disorder, and improved crystalline quality. This result is also consistent with the larger grain sizes and higher open-circuit voltages reported in selenium solar cells from DTU. Finally, Fig.~\ref{fig:PLfigure}(d) compares the PL spectrum with an external quantum efficiency (EQE) measurement from the same lab and its derivative, highlighting a significant difference between the optical and photovoltaic bandgaps. This shift indicates a broad absorption onset, which is detrimental for photovoltaic performance \cite{kirchartz2018a}. While part of this effect may be attributed to an Urbach tail, no systematic studies have yet separated its static and dynamic components, where the static contribution could be reduced by further improving the structural quality \cite{ugur2022a}. Overall, the origin of the shift between PL emission and absorption/collection remains unresolved, and it is unclear whether this gap is intrinsic or can be at least partially reduced through targeted material engineering.

\begin{figure*}[t!]
    \centering
    \includegraphics[width=\textwidth,trim={0 0 0 0},clip]{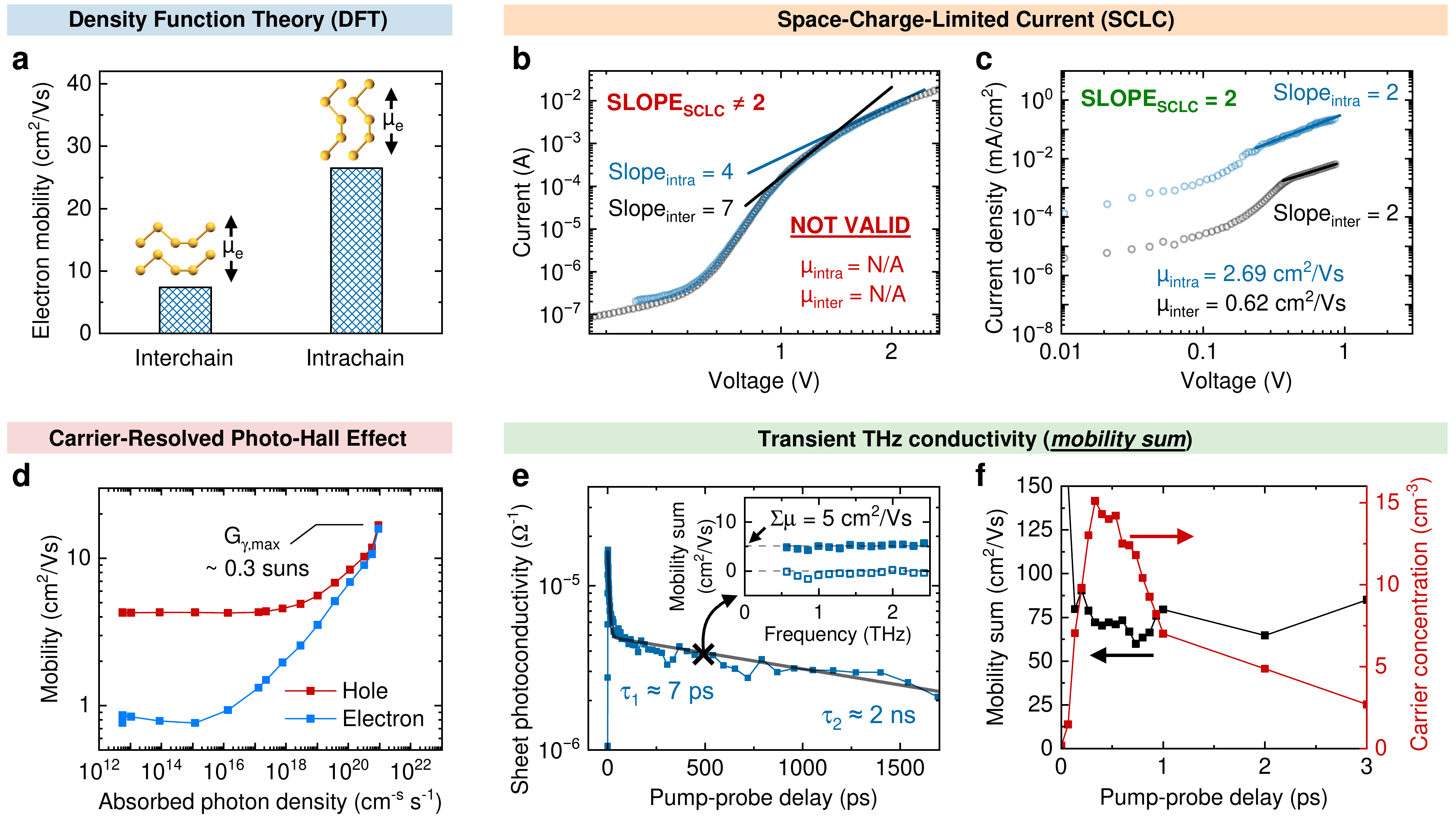}
    \caption{Reported carrier mobilities in trigonal selenium thin films. (a) DFT-calculated electron mobilities along the intra- and interchain directions from Ref.~\cite{liu2025a}. (b, c) Dark IV curves used to extract intra- and interchain electron mobilities from the SCLC region from Refs.~\cite{liu2025a} and \cite{shen2025a}, respectively. (d) Carrier-resolved mobilities as a function of light intensity measured using the photo-Hall effect from Ref.~\cite{nielsen2025a}. (e, f) OPTP measurements showing sheet photoconductivity decay and conductivity spectra fitted using Drude-Smith modeling to retrieve the sum of electron and hole mobilities from Refs.~\cite{nielsen2022a} and \cite{nielsen2023b}, respectively.}
    \label{fig:MobilityFigure}
\end{figure*}

\subsection{Effective DOS and Carrier Masses}\label{sec:DOSandMasses}

The effective density-of-states (DOS) carrier masses strongly influence the photovoltaic performance potential of a photoabsorber due to the inverse quadratic dependence on the quasi-Fermi level splitting, $\mathrm{QFLS} \propto \ln \left( 1/\left[ m_em_h\right]^{3/2}\right)$ \cite{kirchartz2018b, crovetto2024b}. In fact, the open-circuit voltage is more sensitive to the effective DOS carrier masses than to doping. The effective DOS is commonly obtained from band structure calculations by fitting the band-edge dispersion with a parabolic relation. However, as the valence band edge of trigonal selenium exhibits significant non-parabolicity, the effective DOS must be determined through explicit band integration as described in Ref.~\cite{nielsen2022a}. At $T=300 \, \mathrm{K}$, we obtain

\begin{equation}
    N_\mathrm{V} = 1.64 \times 10^{20} \, \mathrm{cm}^{-3} , \quad
    N_\mathrm{C} = 8.7 \times 10^{18} \, \mathrm{cm}^{-3}
\end{equation}

\noindent and from the effective DOS and the optical bandgap determined from PL spectroscopy, the intrinsic carrier concentration at $T=300$~K is

\begin{equation}
    n_\mathrm{i} = \sqrt{N_\mathrm{V}N_\mathrm{C} \exp\left(-\frac{E_\mathrm{g}}{k_\mathrm{B}T}\right)} = 1.09 \times 10^{4}\,\mathrm{cm}^{-3}
\end{equation}

\noindent which is an order of magnitude higher than our previous estimate, as the optical bandgap from PL spectroscopy is $\sim$100~meV lower than that obtained from our Tauc plot extrapolations. The effective DOS masses from Brillouin zone integration are

\begin{equation}
    m_{\mathrm{dv}}^* = 3.50 \times m_0, \qquad m_{\mathrm{dc}}^* = 0.49 \times m_0
\end{equation}

\noindent It is important to note that the conductivity effective masses that are relevant for charge transport are not equivalent to the effective DOS masses. Evaluating the tensor elements based on band curvature and crystal symmetry, the conductivity effective masses are determined to be

\begin{align}
    m_{\mathrm{v}\perp} = 0.68 \times m_0   \qquad    m_{\mathrm{v}\parallel} = 0.21 \times m_0\\
    m_{\mathrm{c}\perp} = 0.39 \times m_0   \qquad    m_{\mathrm{c}\parallel} = 0.18 \times m_0
\end{align}

\noindent and combined with the mean intra-band carrier scattering times, these masses are directly related to the carrier mobilities discussed in the following.

\subsection{Carrier Mobilities}

Carrier mobilities in trigonal selenium span a wide range of values in the literature, from $\mu_\text{TRMC}\approx0.5$~cm$^\text{2}$/Vs measured on powders using pump-probe time-resolved microwave conductivity (TRMC) \cite{bhaskar2017a}, to $\mu_\text{Hall}=260$~cm$^\text{2}$/Vs measured on single crystals using Hall effect measurements \cite{heleskivi1969a}. In 1949, Henkels studied the carrier mobilities and density in selenium single crystals using thermoelectric and resistivity measurements, and reported intrachain and interchain mobilities of 5~cm$^2$/Vs and 1~cm$^\text{2}$/Vs, respectively \cite{henkels1950a}. More consistent values have been reported from single crystal studies using acoustoelectric current satuation \cite{mort1967a}, magnetoconductivity \cite{mell1967a}, and phonon drag \cite{stuke1965a} resulting in hole mobilities of $\mu_\text{h}=28$~cm$^\text{2}$/Vs along the chain direction, with values approximately 3-4 times lower perpendicular to the chains. As shown in Fig.~\ref{fig:MobilityFigure}(a), first-principle calculations by Liu \textit{et al.} predict a comparable anisotropy for electron mobilities, with $\mu_\text{e}=26.5$~cm$^2$/Vs and $7.4$~cm$^2$/Vs along and perpendicular to the chains, respectively \cite{liu2025a}. Finally, I want to highlight a particularly interesting study by Mort \textit{et al.} from 1968, showing that, under sufficiently intense background illumination, transient photoconductivity measurements give higher mobilities of $40\pm8$~cm$^2$/Vs and $17\pm4$~cm$^2$/Vs along and perpendicular to the chains, respectively \cite{mort1968a} -- this result suggests illumination-dependent carrier mobilities. However, these studies focus on single crystals, whereas the effective mobilities in selenium thin films that are relevant for photovoltaic devices may be strongly influenced by strain, grain boundaries, structural disorder, and interfaces, and therefore do not necessarily translate directly from single crystals measurements.

To date, only two groups have reported preferentially oriented trigonal selenium thin films with directionally resolved carrier mobility measurements. In both cases, the mobility was extracted using space-charge-limited current (SCLC) analysis. To directly compare IV curves of the intra- and interchain devices, I digitized and compiled the IV curves from both groups, shown in Fig.~\ref{fig:MobilityFigure}(b) and (c). SCLC-measurements require a device in which only one type of carrier is transported, which in these studies is electrons. In the first report by Liu et al. \cite{liu2025a}, the device structure was FTO/TiO$_\text{2}$/Se/Au, but this is not a single-carrier device as it allows for asymmetric carrier transport. Furthermore, as shown in Fig.~\ref{fig:MobilityFigure}(b), the control device with randomly oriented grains and the substrate-heated device with preferentially oriented grains are nearly indistinguishable. Moreover, the quadratic relation between current and voltage in the theoretical SCLC-regime requires the slope to be 2 for the analysis to be valid, yet the digitized linear fit regions in the double-log plot exhibit slopes of 4 for the intra- and 7 for the interchain devices, respectively. Finally, changes in carrier mobility should produce an absolute offset in current, but the overlapping IV curves show no such offset, despite the reported mobilities differing by a factor of $\approx4$. Given these issues, I consider the results from these measurements unreliable and have therefore labeled the mobilities in Fig.~\ref{fig:MobilityFigure}(b) invalid.

The SCLC-measurements from Shen \textit{et al.} \cite{shen2025a} are shown in Fig.~\ref{fig:MobilityFigure}(c). Here, the slopes of the two linear fits in the SCLC region of the double-log plot are 2, and the intra- and interchain IV curves are offset by an absolute constant, consistent with different electron mobilities. The single-carrier device requirement is also satisfied by the reported electron-only structure, FTO/TiO$_\text{2}$/t-Se/PCBM/Ag. While a reasonable anisotropic mobility ratio between intra- and interchain directions is observed, the absolute values deviate substantially from single crystal hole mobilities and the DFT-calculated electron mobilities. However, I think it's worth highlighting that SCLC analysis as a whole has received extensive critique \cite{roehr2018a}, emphasizing the need to cross-validate the reported mobilities against other techniques, even when all measurement and analysis criteria are nominally met.

In a study from 2025, we used the carrier-resolved photo-Hall effect to determine both minority and majority carrier mobilities in randomly oriented polycrystalline selenium thin films under varying steady-state illumination conditions \cite{nielsen2025a}. The results, shown in Figure \ref{fig:MobilityFigure}(d), indicate that the electron mobility under dark conditions is comparable to the interchain mobility reported in Fig.~\ref{fig:MobilityFigure}(c). As the light intensity increases, the mobilities of both carriers rise, converging toward the same value, implying illumination-dependent mobilities in agreement with the transient photoconductivity study by Mort \textit{et al.} from 1968 \cite{mort1968a}. Although the measurement setup was limited to a maximum illumination level equivalent to $\approx0.3$ suns, the observed trend suggests that the mobilities could increase further, suggesting that the reported values should be regarded as lower bounds for the mobilities in selenium thin films under photovoltaic operating conditions. Furthermore, given that the Hall measurements probe lateral transport across thousands of grain boundaries through randomly oriented crystal, and the relevant mobility for devices corresponds to transport within a single grain along the charge transport direction, the measured values of $\mu_\text{e}=\mu_\text{h}\approx 16$~cm$^\text{2}$/Vs at 0.3 suns represent a conservative estimate of the effective mobilities.

The carrier-resolved photo-Hall mobilities contradict an earlier assessment of the carrier mobilities from our group using optical pump THz probe spectroscopy (OPTP) \cite{nielsen2022a}. In that study, we photoexcited selenium thin films with an optical pump and measured the decay of the sheet photoconductivity. We then fitted the frequency-dependent conductivity spectrum at a pump-probe delay of 0.5~ns with a Drude-Smith model to retrieve the sum of the electron and hole mobilities, $\Sigma \mu = \mu_\text{e}+\mu_\text{h} \approx 5$~cm$^\text{2}$/Vs, as shown in Fig. \ref{fig:MobilityFigure}(e). While this mobility sum agrees reasonably well with the sum of the photo-Hall mobilities under dark or low-injection conditions, we had interpreted the slower of the two characteristic decay times as the effective carrier lifetime. As a consequence, we restricted our mobility assessment to this regime. However, when we compare the photo-Hall data, photoconductivity decay, and mobility sum, I now believe that the initial decay is the result of a rapid localization of the photoexcited carrier. Therefore, to better understand the free carrier dynamics before this initial decay, we repeated the OPTP measurements, revealing a mobility sum of $\Sigma \mu \approx 75$~cm$^\text{2}$/Vs \cite{nielsen2023b} in the first few picoseconds, which is shown in Fig.~\ref{fig:MobilityFigure}(f). Collectively, these results indicate that our first interpretation of the photoconductivity decay times was likely incorrect, and that we need to reassess what the effective carrier lifetime in state-of-the-art selenium thin films actually is.

\begin{figure*}[t!]
    \centering
    \includegraphics[width=\textwidth,trim={0 0 0 0},clip]{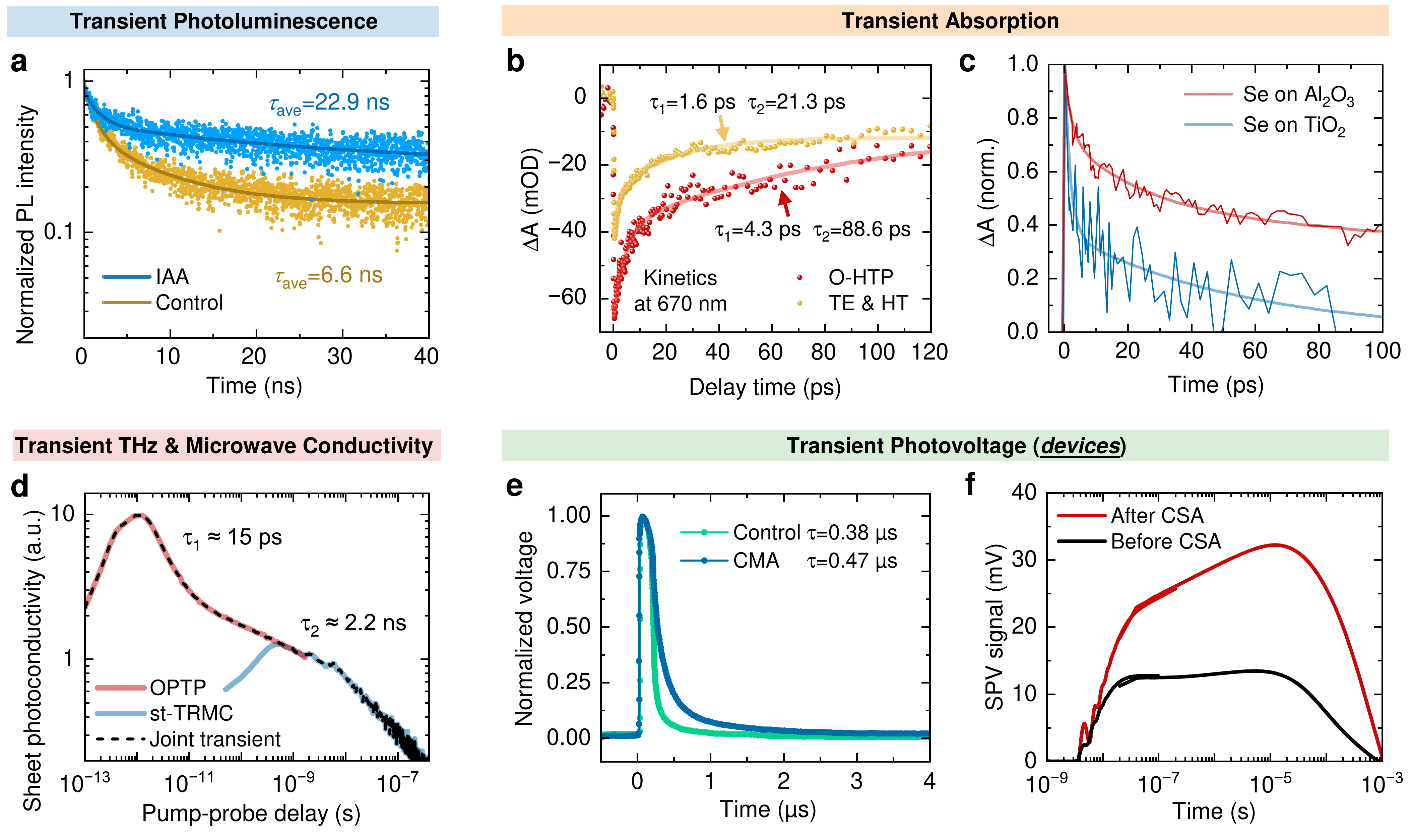}
    \caption{Reported carrier lifetimes in trigonal selenium thin films. (a) Transient photoluminescence measurements fitted with bi-exponential decay functions from Ref.~\cite{wen2026a}, where $\tau_\text{ave}$ is the weighted average decay times. (b, c) Transient absorption spectroscopy showing picosecond carrier dynamics from Refs.~\cite{shen2025a} and \cite{zhu2016a}, respectively. (d) Joint transient from optically pumped THz probe (OPTP) and transient microwave conductivity (TRMC) measurements illustrating carrier dynamics across pico- and nanosecond timescales from Ref.~\cite{nielsen2024b}. (e, f) Transient photovoltage and surface photovoltage (SPV) measurements capturing the lifetime of spatially separated charge carriers in photovoltaic devices from Refs.~\cite{lu2024a} and \cite{nielsen2024b}, respectively.}
    \label{fig:LifetimeFigure}
\end{figure*}

\subsection{Carrier Lifetimes}\label{sec:Lifetimes}

Time-resolved photoluminescence (TRPL) is arguably the most commonly used technique to determine carrier lifetimes. However, likely due to the difficulty of measuring PL even in state-of-the-art samples, only two studies have reported TRPL transients for selenium thin films. The first, by Liu \textit{et al.} \cite{liu2023a}, is associated with the steady-state PL spectrum labeled III in Fig.~\ref{fig:PLfigure}(a), which features two radiative emission bands, neither of which is likely to originate from band-to-band recombination. The fitted mono-exponential decay time of $\tau_1=129$~ns is therefore more reasonably attributed to carriers trapped in long-lived defect states. The second study by Wen \textit{et al.} \cite{wen2026a}, shown in Fig.~\ref{fig:LifetimeFigure}(a), reports a weighted average lifetime of $\tau_{\mathrm{ave}}=22.9$~ns for the best-performing sample. While this result is remarkable, it should be assessed with caution. First, the low signal-to-noise ratio suggests an overall weak PL signal and low PL quantum yield. Second, the pre-excitation ($t<0$) background signal is not shown, and the decay extends beyond the recorded 40~ns window, indicating that the system has not fully relaxed within the measurement range. Finally, the reported lifetime is three orders of magnitude longer than that of most other reports for selenium thin films, yet corresponding devices exhibit a comparable open-circuit voltage deficit. This raises questions about potential issues with the experimental procedures, such as insufficient filtering of the excitation source allowing for the tail of the laser to bleed into the detector unit, which cannot be excluded based on the reported methodology. If valid, however, the reported carrier lifetime would exceed those of many emerging PV absorbers and re-open the question of the origin of the still-significant open-circuit voltage losses.

\begin{figure*}[t!]
    \centering
    \includegraphics[width=\textwidth,trim={0 0 0 0},clip]{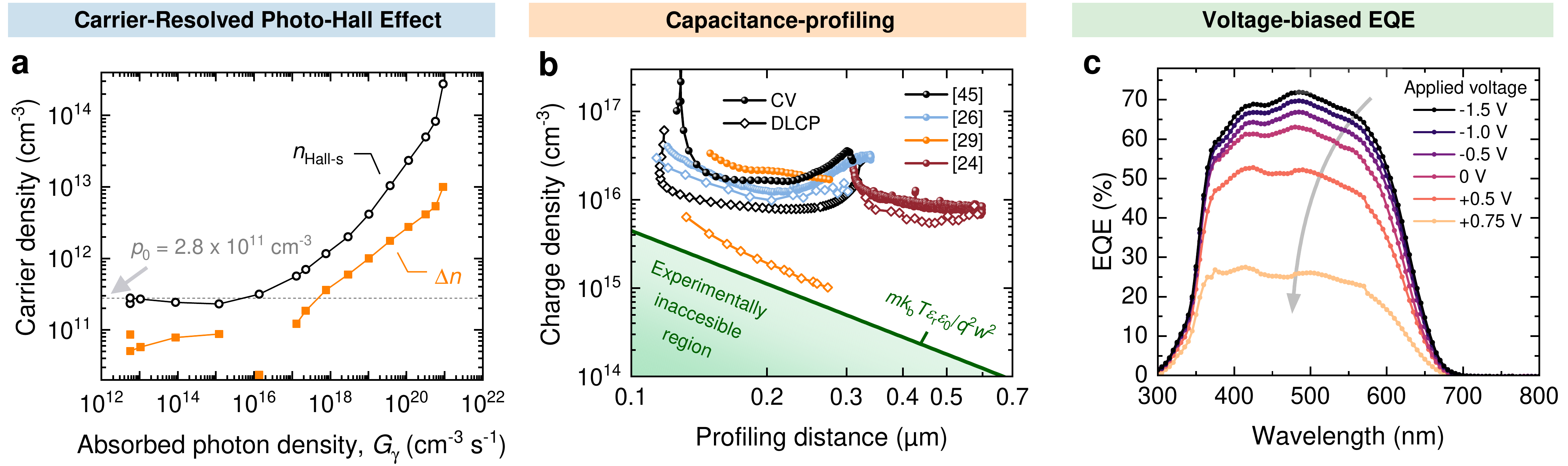}
    \caption{Reported doping density in trigonal selenium thin films. (a) Carrier-resolved photo-Hall measurements from Ref.~\cite{nielsen2025a} showing the extracted carrier density as a function of absorbed photon density. (b) Digitized and compiled charge density profiles derived from capacitance-voltage (CV) and drive-level capacitance profiling (DLCP) measurements from Refs.~\cite{fu2022a, zheng2022a, yan2022a, nielsen2025a}. The experimentally inaccessible region is highlighted in green, reflecting the fundamental limitation imposed by geometrical capacitance and charge injection from the carrier-selective contacts. (c) EQE spectra of state-of-the-art selenium solar cells under varying applied voltage bias from Ref.~\cite{nielsen2024b}.}
    \label{fig:DopingFigure}
\end{figure*}

Moving into the picosecond regime, the temporal resolution of typical PL detectors ($\approx300$~ps \cite{hempel2018a}) is insufficient to resolve the relevant carrier dynamics. This may partly explain the lack of TRPL measurements in the selenium literature, if the effective carrier lifetime is on the picosecond timescale. In the work by Shen \textit{et al.} \cite{shen2025a}, featuring selenium thin films with and without preferentially oriented crystal grains, transient absorption spectroscopy (TAS) measurements were presented with pump-probe delays up to 120 ps post-excitation. Two characteristic decay components are extracted from the transient signals shown in Fig.~\ref{fig:LifetimeFigure}(b), although the system does not reach a fully relaxed state within this time window. Given that the synthesis strategies and device performances are broadly comparable between the works of Shen \textit{et al.} and Liu \textit{et al.}, it is puzzling that one study reports picosecond carrier dynamics while the other reports nanosecond lifetimes. Earlier work by Zhu \textit{et al.} \cite{zhu2016a}, shown in Fig.~\ref{fig:LifetimeFigure}(c), also presented transient absorption spectra and kinetics, revealing ultrafast carrier dynamics on the picosecond scale alongside a longer-lived component in the tens of nanoseconds (not shown here) for selenium thin films grown on Al$_\text{2}$O$_\text{3}$. This secondary component, likely associated with long-lived trap states, is comparable to the decay times observed in the TRPL transients in Fig.~\ref{fig:LifetimeFigure}(a), suggesting a potentially common physical origin. In a recent study from our group, we joined the transients decays of optical pump THz probe (OPTP) with time-resolved microwave conductivity (TRMC) measurements to capture carrier dynamics across multiple timescales in the same selenium thin film sample. The resulting transient, shown in Fig.~\ref{fig:LifetimeFigure}(d), confirms the coexistence of ultrafast and longer-lived dynamic processes spanning several orders of magnitude in pump-probe delay time.

Significantly longer apparent lifetimes have been reported using transient photovoltage measurements, but it is important to note that these measurements are performed on selenium thin films contacted with carrier-selective transport layers. The example shown in Fig.~\ref{fig:LifetimeFigure}(e), digitized from the work by Lu \textit{et al.}, exhibits photovoltage decay times on the order of hundreds of nanoseconds \cite{lu2024a}. Similarly, Liu \textit{et al.} reported photovoltage decay times extending into the tens of microseconds \cite{liu2023a}. While these results are valuable, the extracted decay times should not be interpreted as bulk carrier lifetimes. Instead, the photovoltage decay corresponds to the lifetime of spatially separated charge carriers that have already been transported into carrier-selective contacts, where recombination rates can be several orders of magnitude lower due to negligible minority carrier concentrations. To illustrate this distinction, the selenium thin films analyzed using OPTP and TRMC in Fig.~\ref{fig:LifetimeFigure}(d) were processed in parallel with the selenium solar cells studied using transient surface photovoltage measurements, shown in Fig.~\ref{fig:LifetimeFigure}(f). While the carrier dynamics in the bare absorber occur on picosecond to nanosecond timescales, the lifetime of spatially separated carriers in the device structure may be up to milliseconds.

Collectively, the mobilities and transient conductivities shown in Figs.~\ref{fig:MobilityFigure} and \ref{fig:LifetimeFigure} suggest ultrafast free-carrier dynamics followed by localization into a longer-lived trap states. To test this hypothesis, one could design a pump-probe-push experiment. The initial pump pulse photoexcites the system, while the probe monitors the excitation and subsequent rapid decay. Before the system has fully relaxed, a lower-energy push pulse is introduced to promote the localized carriers back into delocalized states. If the hypothesis is correct, the energy of the push pulse should be sufficient re-excite these carriers and transiently restore high mobilities. The assumption of highly mobile yet ultrashort-lived free carriers will be revisited in Section~\ref{sec:SCAPS} using drift-diffusion simulations to assess the implications for device-level losses.

\subsection{Doping Density} 

The doping density in trigonal selenium thin films is widely debated, with reported values spanning several orders of magnitude. Experimentally, however, the material is consistently found to be p-type. This doping polarity is consistent with the DFT-predicted dopability window calculated by Kavanagh \textit{et al.}, where vacancy-induced charge compensation prevents n-type doping (see Fig.~\ref{fig:IntrinsicDefects}) \cite{kavanagh2025a}. In the same study, a wide range of extrinsic point defects and their influence on the hole concentration were evaluated. Both intrinsic and extrinsic defects were found to contribute negligibly to doping, either due to electrical inactivity (chalcogens) or self-compensation (hydrogen, halogens, pnictogens). Halogen impurities were identified in the bulk of state-of-the-art selenium thin films using time-of-flight secondary ion mass spectrometry, but even when including high concentrations of fluorine and chlorine in the DFT calculations, the predicted doping density remains limited to $p\sim10^{12}$~cm$^{-3}$. This value contrasts sharply with the more commonly reported experimental values of $p\sim10^{16}$~cm$^{-3}$.

Hall effect measurements are a classical workhorse in semiconductor characterization used to determine the type, density, and mobility of the majority carriers. For trigonal selenium single crystals, Heleskivi \textit{et al.} reported a hole concentration of $p=0.5\times10^{15}$~cm$^{-3}$ using DC Hall measurements in 1969 \cite{heleskivi1969a}. In contrast, Todorov \textit{et al.} reported a much lower carrier density of $p=2.8\times10^{12}$~cm$^{-3}$ using parallel dipole line AC Hall measurements \cite{todorov2017a}. However, this measurement was performed on a 105~nm selenium thin film deposited on ZnMgO, a more highly doped n-type semiconductor. Therefore, it would be reasonable to suspect that the selenium thin film is fully depleted. This could explain the large discrepancy between the reported carrier densities. This motivated a collaborative study using the carrier-resolved photo-Hall method developed by Gunawan \textit{et al.} \cite{gunawan2019a} to extract both majority and minority carrier properties in state-of-the-art selenium thin films grown directly on quartz substrates \cite{nielsen2025a}. Despite eliminating substrate-induced depletion, the measured carrier densities under dark and low-illumination conditions remained on the order of $p\sim10^{12}$~cm$^{-3}$, as shown in Fig.~\ref{fig:DopingFigure}(a). However, complementary temperature-dependent Hall measurements revealed that carrier freeze-out and depletion from surface and interface defects strongly influenced our results, implying that the measured carrier density does not necessarily reflect the true acceptor density. For Se$_\text{1-x}$Te$_\text{x}$ alloys, Zheng \textit{et al.} reported decreasing carrier densities with decreasing tellurium content \cite{zheng2022a}, reaching $p=10^{14}$~cm$^{-3}$ for Se$_\text{70}$Te$_\text{30}$. However, their Hall measurement setup was not sufficiently sensitive to resolve carrier densities in their elemental selenium thin film.

In addition to Hall effect measurements, capacitance-voltage (CV) and drive-level capacitance profiling (DLCP) have been widely used to assess the apparent charge density in selenium absorbers as a function of profiling depth in completed solar cells \cite{zheng2022a, nielsen2022a, chen2024a, yan2022a, nielsen2023a, fu2022a}. These profiles, digitized and compiled in Fig.~\ref{fig:DopingFigure}(b), show that CV measurements across all the independent studies consistently yield $p\sim10^{16}$~cm$^{-3}$, in agreement with DC Hall measurements on large single crystals. However, CV-based techniques have a fundamental lower detection limit set by the geometrical capacitance and charge injection from carrier-selective contacts \cite{ravishankar2021a, ravishankar2022a}. This limitation is particularly relevant for the DLCP profile reported by Yan \textit{et al.}, which approaches the experimentally inaccessible regime. As this result deviates signficantly from both their corresponding CV profile and those reported by other groups, it should be interpreted with caution. In my opinion, a more convincing validation of the CV- and DLCP-derived doping densities would require a systematic study of the evolution of the charge density plateau with significantly thicker selenium absorber layers.

Hall effect measurements on single crystals yield $p\sim10^{15}$~cm$^{-3}$, while values on the order of $p\sim10^{12}$~cm$^{-3}$ are typically found for thin films. In contrast, capacitance profiling generally gives $p\sim10^{16}$~cm$^{-3}$. An intermediate value of $p\sim10^{14}$~cm$^{-3}$ was reported by Henkels in 1949 for trigonal selenium single crystals, derived from combined thermoelectric and temperature-dependent resistivity measurements \cite{henkels1950a}. What makes this study particularly interesting is that the same methodology applied to microcrystalline samples revealed a strong dependence on crystal size, with smaller crystallites showing higher carrier concentrations and a correspondingly strong sensitivity to the crystallization temperature.

Another key observation is the voltage-bias dependence of the EQE spectra in state-of-the-art selenium solar cells, shown in Fig.~\ref{fig:DopingFigure}(c) \cite{nielsen2024b}. Under forward bias, the depletion region width decreases, leading to a clear reduction in carrier collection efficiency, while the opposite is observed under reverse bias. This implies that the 300~nm selenium absorber film is not fully depleted under short-circuit conditions. If the true doping density were on the order of $p\sim10^{12}$~cm$^{-3}$, as suggested by Hall effect measurements, the depletion region width would extend over tens of micrometers. By contrast, a doping density of $p\sim10^{16}$~cm$^{-3}$ would deplete roughly 200–250~nm of the absorber, which is more consistent with the observed EQE bias dependence. These observations support the idea that Hall measurements are strongly influenced by depletion from surface and interface defects, potentially combined with carrier freeze-out at room temperature. Conversely, if we accept the low doping density predicted by DFT and inferred from Hall measurements, we need an alternative explanation for both the electrical bias-dependence of the EQE-spectra and the persistent charge density plateaus reported across multiple capacitance profiling studies. An unambigious experimental determination of the doping density has yet to de demonstrated, and no computational framework currently explains doping levels above $p>10^{12}$~cm$^\text{-3}$. In Section~\ref{sec:SCAPS}, we turn to drift-diffusion simulations to aid the determination of this key material parameter, which strongly influences the photovoltaic performance potential of selenium solar cells.

\begin{figure*}[t!]
    \centering
    \includegraphics[width=\textwidth,trim={0 0 0 0},clip]{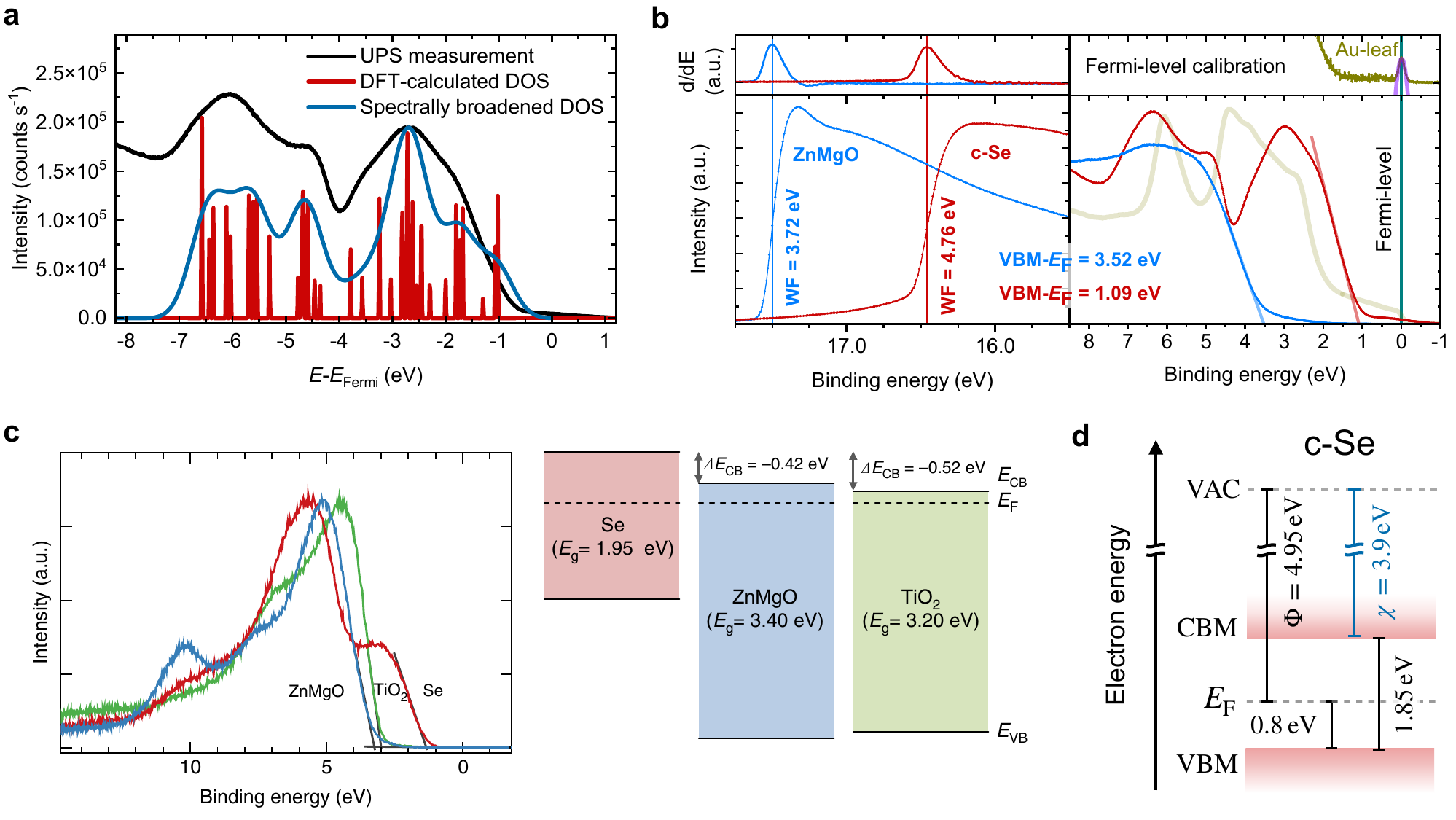}
    \caption{Electronic band positions in trigonal selenium thin films. (a) UPS measurements compared with the spectrally broadened DFT-calculated valence band structure from Ref.~\cite{nielsen2022a}. (b) UPS measurements of selenium and ZnMgO thin films from Ref.~\cite{nielsen2024a}, where a small bias of –5~V was applied to the samples to deconvolute the work function of the analyzer from the true work function of the sample surface. A gold leaf was deposited on both samples to calibrate the binding energy scale to the Fermi-level of a metal in electrical contact with the sample surface. 
    (c) Femtosecond pump-probe UPS spectra of selenium, TiO$_\text{2}$, and ZnMgO from Ref.~\cite{todorov2017a}, where the optical pump is used to flatten the energy bands at the surface. The corresponding conduction band alignment is constructed by adding the optical bandgap extracted from Tauc plots. (d) Updated absolute energy band positions of trigonal selenium thin films compiled from literature reports and assessed for validity.}
    \label{fig:BandAlignmentFigure}
\end{figure*}

\subsection{Energy Band Positions}\label{sec:BandPositions}

Even though the performance potential of a photovoltaic absorber material does not explicitly depend on the absolute positions of the band edges -- hence their absence from the FOM in Eq.~\ref{eq:FOM} -- these positions are critical for identifying suitable carrier-selective contact materials. The most common approach in determining band positions is to extract the valence band maximum (VBM) using photoemission spectroscopy (XPS or UPS), and subsequently add the optical bandgap to arrive at the conduction band minimum (CBM). In this framework, the VBM and CBM correspond to the ionization energy and electron affinity, respectively. However, this methodology is inherently prone to significant uncertainties. First, the CBM is derived by combining two independent measurements, each introducing its own errors. Second, photoemission spectroscopy is intrinsically surface-sensitive, making it susceptible to band bending from surface defects, surface photovoltage effects induced by the probe, and Fermi-level pinning. Moreover, the binding energy scale is often not calibrated using a metallic reference in electrical contact with the sample surface, which is required to ensure a common Fermi level. While the absorber should ideally be grown on a device-relevant substrate, the issue is further exacerbated if the film is insufficiently thick, as substrate-induced depletion can shift the Fermi level and the distort the true unaffected VBM of the bulk. This effect has been clearly demonstrated by Schulz \textit{et al.} \cite{schulz2015a}, where the apparent doping polarity of halide perovskites flips from n-type when grown on TiO$_\text{2}$ to p-type when grown on NiO$_\text{x}$, corresponding to a dramatic shift of the VBM of up to 0.7~eV. Finally, insufficient reporting of acquisition conditions in many studies makes it difficult to assess the reliability and uncertainty of the reported band positions. For a comprehensive discussion of minimum reporting requirements and common pitfalls, I refer to a great work on best practices in PV materials characterization by Hoye \textit{et al.} \cite{hoye2017a}.

\begin{figure*}[t!]
    \centering
    \includegraphics[width=\textwidth,trim={0 0 0 0},clip]{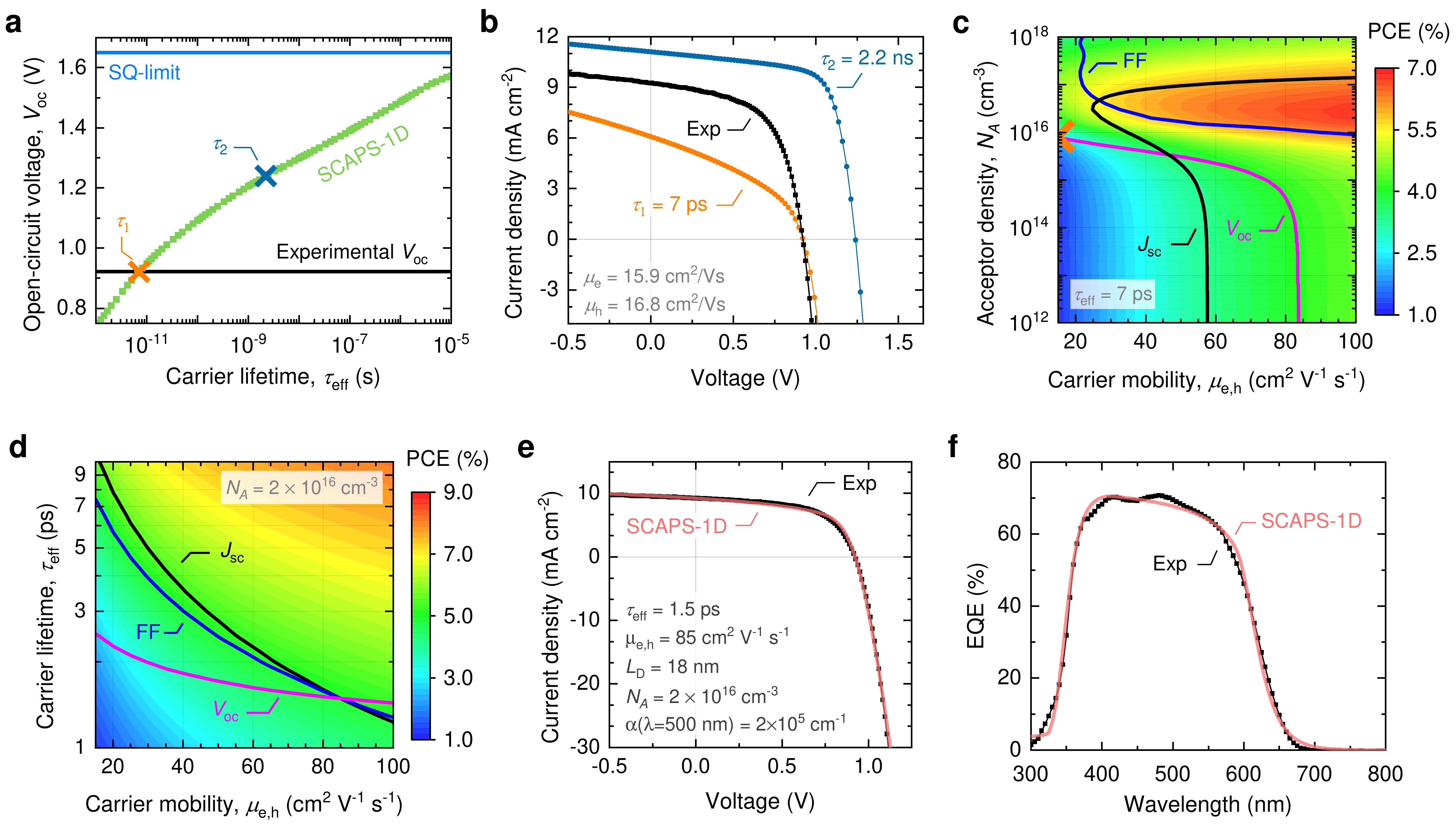}
    \caption{Drift-diffusion simulations of selenium solar cells. (a) Simulated open-circuit voltage as a function of effective carrier lifetime. Here, $\tau_\text{1}$ and $\tau_\text{2}$ represents the two characteristic decay times from Fig.~\ref{fig:MobilityFigure}(e). (b) Experimental and simulated JV-curves assuming these two characteristic lifetimes. (c) Contour plot of the power conversion efficiency (PCE) when varying the acceptor density and carrier mobilities in tandem, assuming an effective carrier lifetime of $\tau = 7$~ps. The solid lines represents a match between the simulated and experimental JV parameters. (d) Contour plot of the PCE when varying the carrier lifetime and mobilities in tandem, assuming an acceptor density of $N_A=2\times10^{16}$~cm$^{-3}$. (e, f) JV-curve and EQE spectrum of the best simulated match compared with the experimental results. The material properties of the absorber are summarized in (e). Adapted with permission from Ref.~\cite{nielsen2025a}.}
    \label{fig:SCAPSFigure}
\end{figure*}

The VBM of trigonal selenium thin films has been reported to lie 0.74~eV below the Fermi level by Bao \textit{et al.} \cite{bao2025a}, which is in good agreement with our earlier result from 2021 \cite{nielsen2021a}. When combined with the optical bandgap determined from PL spectroscopy, this position is consistent with p-type doping, in line with DFT, Hall measurements, capacitance-based techniques, and the diode behavior observed when pairing selenium with n-type semiconductors. However, in both studies, the binding energy scale was not reported to be calibrated using a metallic reference in electrical contact with the sample surface, introducing significant uncertainty. To address this, we later deposited a gold leaf in electrical contact with the selenium film to properly calibrate the binding energy scale. This resulted in a VBM position more than 1~eV below the Fermi level, as shown in Fig.~\ref{fig:BandAlignmentFigure}(b), which counterintuitively suggests n-type doping. A similar n-type Fermi level position is observed in the work by Todorov \textit{et al.} \cite{todorov2017a}, which is shown in Fig.~\ref{fig:BandAlignmentFigure}(c). Here, the authors used femtosecond pump–probe UPS to mitigate surface band bending and effectively flatten the bands \cite{lim2005a}. However, as the absorber is optically pumped, the Fermi level is split into two quasi-Fermi levels. Given that the electron quasi-Fermi level is expected to lie closer to the CBM, the absolute band positions become difficult to interpret in a physically meaningful way.

In addition to determining the VBM relative to the Fermi level, we applied a small electrical bias to the sample surface to deconvolute the work function of the electron analyzer from the true work function of the surface. From this, we obtained a work function of 4.76~eV, corresponding to a VBM position of 5.84~eV relative to the vacuum level. Using photoemission yield spectroscopy in air, Hadar \textit{et al.} reported VBM positions for Se$_\text{1-x}$Te$_\text{x}$ alloys, where a value of 5.67~eV was measured for the pure selenium sample \cite{hadar2019a}.

For the determination of the CBM, our previous studies, as well as the work by Todorov \textit{et al.} \cite{todorov2017a}, used an optical bandgap of 1.95~eV extracted from Tauc plots. Now, having concluded that the optical bandgap is closer to 1.85~eV from PL spectroscopy, the electron affinity must be reassessed. Using the previously determined VBM position of 5.84~eV, this gives a CBM position of 3.99~eV relative to the vacuum level. While this is slightly higher than the previously adopted value of 3.89~eV, it remains within the uncertainties associated with the methodology. Hadar \textit{et al.} also determined the CBM of the aforementioned Se$_\text{1-x}$Te$_\text{x}$ alloys using optical bandgaps extracted from Tauc plots, reporting a CBM position of 3.88~eV for pure trigonal selenium in the main manuscript of Ref.~\cite{hadar2019a}. However, there is an inconsistency between the main text and the supplementary information, where a value of 3.94~eV is given for the pure selenium sample.

In summary, the VBM position relative to vacuum has been reported between 5.84 and 5.67~eV, and the CBM between 3.99 and 3.88~eV. Combining these results across several independent studies, with the revised optical bandgap of 1.85~eV from PL spectroscopy, I have updated my best qualified estimate of the band positions for trigonal selenium in the schematic shown in Fig.~\ref{fig:BandAlignmentFigure}(d).

\subsection{Drift-Diffusion Simulations}\label{sec:SCAPS}

\begin{table*}[ht!]
\small
  \caption{\ Material parameters used in SCAPS-1D device simulations from Ref \cite{nielsen2025a}.}
  \vspace{0.25cm}
  \label{tbl:SCAPSparameters}
  \begin{tabular*}{\textwidth}{@{\extracolsep{\fill}}lrrrr}
    \hline \vspace{-0.25cm}\\
    \textbf{Contact properties} & \textbf{Front} & & & \textbf{Back} \\
    $S_\mathrm{e}$ (cm/s) & $\text{10}^{\text{7}}$ & & & $\text{10}^{\text{7}}$ \\
    $S_\mathrm{h}$ (cm/s) & $\text{10}^{\text{7}}$ & & & $\text{10}^{\text{7}}$ \\
    $\phi$ (eV) & 4.32 (flat bands) & & & 5.20 \\
    $T/R$ (\%)\vspace{0.2cm} & 80/0 & & & 0/100 \\
    \hline \vspace{-0.25cm}\\
    \textbf{Layer properties}$^\star$ & \textbf{FTO} & \textbf{ZnMgO} & \textbf{poly-Se}$^*$ & \textbf{MoO$_\text{x}$} \cite{li2019a} \\
    Thickness (nm) & $\text{500}$ & $\text{65}$ & $\text{300}$ & $\text{15}$ \\
    $E_\mathrm{g}$ (eV) & 3.80 \cite{banyamin2014a} & $\text{3.55}$ \cite{nielsen2024a} & $\text{1.95}$ \cite{nielsen2022a} & 3.00 \\
    $\chi$ (eV) & 4.44 & 3.69 \cite{nielsen2024a} & 3.89 \cite{nielsen2024a} & 2.50 \\
    $\epsilon_\mathrm{r}$ & 9.0 \cite{zyoud2021a} & 8.0 \cite{ashkenov2003a} & 7.8 \cite{nielsen2022a} & 12.5 \\
    $N_C$ (cm$^{-3}$) & $\text{1.00}\times \text{10}^{\text{18}}$ \cite{zyoud2021a} & $\text{2.77}\times \text{10}^{\text{18}}$ \cite{franz2013a} & $\text{8.70}\times \text{10}^{\text{18}}$ \cite{nielsen2022a} & $\text{2.20}\times \text{10}^{\text{18}}$ \\
    $N_V$ (cm$^{-3}$) & $\text{1.00}\times \text{10}^{\text{18}}$ \cite{zyoud2021a} & $\text{3.23}\times \text{10}^{\text{18}}$ \cite{franz2013a} & $\text{1.64}\times \text{10}^{\text{20}}$ \cite{nielsen2022a} & $\text{1.80}\times \text{10}^{\text{19}}$ \\
    $\mu_\mathrm{e}$ (cm$^{2}$/Vs) & 22 \cite{luangchaisri2012a} & 1.5 \cite{jamarkattel2022a} & 85$^\ddag$ & 25 \\
    $\mu_\mathrm{h}$ (cm$^{2}$/Vs) & 10 \cite{zyoud2021a} & 1.5$^\dag$ & 85$^\ddag$ & 100 \\
    $N_{D/A}$ (cm$^{-3}$)\vspace{0.2cm} & $N_D= \text{8.1} \times \text{10}^{\text{20}}$ & $N_D= \text{3.2} \times \text{10}^{\text{18}}$ \cite{jamarkattel2022a} & $N_A= \text{2} \times \text{10}^{\text{16}}$ & $N_A= \text{6.0} \times \text{10}^{\text{18}}$ \\
    \hline \vspace{-0.25cm}\\
    \textbf{Defect states} & \textbf{FTO} & \textbf{ZnMgO} & \textbf{poly-Se}$^\mathsection$ & \textbf{MoO$_\text{x}$} \\
    Type & - & - & Neutral & - \\
    Energy distribution & - & - & Single level & - \\
    $N_\mathrm{t}$ (cm$^{-3}$) & - & - & $\text{1.5} \times \text{10}^{\text{16}}$ & - \\
    $E_\mathrm{t}$ (eV) & - & - & $\textit{E}_\text{V} + \text{0.6}$ & - \\
    $\sigma_\mathrm{e}$, $\sigma_\mathrm{h}$ (cm$^{2}$)\vspace{0.2cm} & - & - & $\text{4.5} \times \text{10}^{-\text{12}}$ & - \\
    \hline \vspace{-0.25cm}\\
    \textbf{Interface defect} & \textbf{FTO/ZnMgO} & \textbf{ZnMgO/Se} & \textbf{poly-Se/MoO$_\text{x}$} & \\
    Type & - & - & - & \\
    Energy distribution & - & - & - & \\
    $E_\mathrm{t}$ (eV) & - & - & - & \\
    $N_\mathrm{t}$ (cm$^{-2}$) & - & - & - & \\
    $\sigma_\mathrm{e}$, $\sigma_\mathrm{h}$ (cm$^{2}$) & - & - & - & \\
    Intraband tunneling & Enabled & Enabled & - & \\
    \qquad \vspace{-0.25cm} \\
    \hline \vspace{-0.25cm}\\
  \end{tabular*}
  \begin{minipage}{\textwidth}
    \raggedright
    \small $^\star$ \footnotesize The thermal velocity of electrons and holes in all layers are assumed to be $\nu_\mathrm{th,(e,h)}=10^7$ cm/s.\\
    \small $^*$ \footnotesize The material properties of selenium, where referenced, are from our previously published works and therefore accurately represent the optoelectronic quality of selenium thin-films studied in this work \cite{nielsen2022a}, synthesized using the same custom tools and in-house recipes. \\ 
    \small $^\dag$ \footnotesize The minority carrier (hole) mobility is assumed to be similar to the majority carrier (electron) mobility in ZnMgO as determined by dark Hall-effect measurements \cite{jamarkattel2022a}.\\
    $^\ddag$ The carrier mobilities in selenium are assumed to be similar under illuminated conditions given the convergence of the Hall-mobility difference $\Delta \mu_\text{Hall} \rightarrow 0$ with increasing absorbed photon density $G_\gamma$.\\
    \small $^\mathsection$ \footnotesize The effective carrier lifetime in selenium is adjusted by varying the capture cross sections, assumed to be similar for both majority and minority carriers, while maintaining a fixed defect density.
  \end{minipage}
  \vspace{-0.4cm}
\end{table*}

It should be clear by now that many material properties are difficult to determine reliably, often leading to conflicting reports in the literature that can misguide research priorities. In this context, numerical device simulations can be a valuable complementary aid in the analysis of material properties and non-idealities. Figure~\ref{fig:SCAPSFigure} shows an example from a recent work of ours, where drift–diffusion modeling in SCAPS-1D was used to systematically explore physically reasonable ranges of material parameters for selenium solar cells, where comparisons across different experimental characterization techniques or with studies from other independent groups did not paint a consistent picture \cite{nielsen2025a}.

The most important objective of such simulations should be to first reproduce experimental results -- in this case, the JV curve and EQE spectrum of a complete device with the structure FTO/ZnMgO/Se/MoO$_\text{x}$/Au. The material parameters for all layers are based on experimental data, computational results, or a combination of both, and are summarized in Table~\ref{tbl:SCAPSparameters}. Given the wide range of reported values for the carrier mobility, doping density, and effective lifetime in selenium thin films, the analysis was first focused on reproducing the experimental open-circuit voltage, which is less sensitive to carrier mobilities as the net charge transport is zero. As discussed in Section~\ref{sec:Lifetimes}, transient absorption and conductivity measurements reveal two characteristic decay times, one in the nanosecond and one in the picosecond regime. Varying the effective carrier lifetime within this range shows that only picosecond lifetimes reproduce the experimentally observed open-circuit voltage.

Adopting a carrier lifetime in the picosecond regime leads to a simulated fill factor (FF) and short-circuit current density (J$_\text{sc}$) values that fall short of the experimental results, as shown in Fig.~\ref{fig:SCAPSFigure}(b). However, the photo-Hall mobilities at 0.3~suns represent a lower bound, which motivates varying the doping density and carrier mobilities in tandem. This is illustrated in the contour plot in Fig.~\ref{fig:SCAPSFigure}(c), where the solid lines represent a match between simulated and experimental JV parameters. Under these assumptions, matching the experimental FF requires an acceptor density of $N_A>10^{16}$~cm$^{-3}$, largely independent of the carrier mobilities. In a second step, the acceptor density is fixed at $N_A=2\times10^{16}$~cm$^{-3}$, while the carrier mobilities and lifetime are varied in tandem, as shown in Fig.~\ref{fig:SCAPSFigure}(d). This simulation reveals a broad range of mobility–lifetime products that reproduce the experimental device performance, provided that the carrier lifetime is constrained to the picosecond regime.

Collectively, these results suggest that the carrier mobilities in selenium thin films obtained under dark and low-injection conditions should be regarded as lower bounds, while the free carrier lifetime is likely governed by an ultrafast localization process on the picosecond scale. The simulations further indicate an acceptor density on the order of $10^{16}$~cm$^{-3}$, significantly exceeding DFT predictions even when extrinsic impurities are considered. As discussed above, the origin of this discrepancy remains unresolved. Finally, the bandgap of selenium was assumed to be 1.95~eV in these simulations, and should be revised to 1.85~eV in future device modeling.

In 2025, Salem \textit{et al.} also used experimental data to establish a baseline in SCAPS-1D \cite{salem2025a}. In this work, the champion device from Todorov et al. \cite{todorov2017a} was simulated to reproduce both the JV curve and EQE spectrum, with the goal of exploring the potential of tandem solar cells combining selenium as the wide-bandgap absorber and antimony selenosulfide (Sb$_\text{2}$(S,Se)$_\text{3}$) as the lower-bandgap absorber. In 2021, Youngman \textit{et al.} simulated the effect of front- versus back-side illumination in bifacial selenium solar cells, predicting the experimentally observed optimal thickness of 300~nm due to limited carrier diffusion lengths \cite{youngman2021a}. While no explicit curve-fitting was applied, the SCAPS-1D simulations captured all experimental trends qualitatively.

The electron-selective transport layer ZnMgO was first introduced in the context of selenium solar cells by Todorov \textit{et al.} in 2017 \cite{todorov2017a}, which significantly improved the open-circuit voltage. The impact of the electron band alignment between ZnMgO and selenium has since been explored through drift–diffusion simulations by Fujimura \textit{et al.} in 2021 \cite{fujimura2022a}, and again by our group in 2024 in the context of selenium/silicon tandem solar cells \cite{nielsen2024a}. In the latter study, SCAPS-1D simulations identified a critical charge transport barrier in the device, which guided experimental efforts resulting in a tenfold improvement in the device efficiency of the tandem solar cell.

Other notable studies using SCAPS-1D simulations of selenium solar cells include Qaid \textit{et al.} from 2023 \cite{qaid2023a}, who also fitted the JV curve and EQE spectrum from Todorov \textit{et al.} to establish a baseline model and subsequently investigated device performance under LED illumination, relevant for indoor PV applications. In 2025, Singh \textit{et al.} \cite{singh2025a} simulated their experimental devices and explored simulation-guided strategies to further increase their device efficiency, as well as the potential of integrating their in tandem with silicon. In 2024, Kobayashi \textit{et al.} \cite{kobayashi2024a} used SCAPS-1D to study the effect of stacked precursors, specifically the influence of moving an ultra-thin tellurium layer a few nanometers into the bulk of the amorphous selenium absorber prior to crystallization. In 2025, He \textit{et al.} \cite{he2025a} investigated the role of interface defects and passivation in selenium solar cells, while Huang \textit{et al.} \cite{huang2025a} studied the impact of different electron-selective transport layers and bulk defect densities under indoor LED illumination.

\section{Thin Film Synthesis and Processing}

The benchmark studies from 1982, 1984, and 1985 by Nakada and Kunioka \cite{kunioka1982a, nakada1984a, nakada1985a} established a baseline for depositing and crystallizing selenium thin films that is still the most commonly used fabrication route today. These works also introduced the most widely used device architecture, featuring an ultra-thin tellurium adhesion layer and TiO$_\text{2}$ as the electron-selective contact. Despite the renewed interest in selenium thin film solar cells, the synthesis and processing steps used to fabricate record-performing devices remain remarkably similar to these early studies. However, subtle variations in processing conditions, along with emerging alternative synthesis pathways, deserve to be examined carefully, as they provide important insight into the crystal growth kinetics of trigonal selenium. This section therefore reviews the various deposition techniques being explored, recent advances in controlling film texture, the role of substrate interactions and surface wetting, strategies for crystallizing amorphous selenium films, and relevant post-processing treatments.

\subsection{Deposition Techniques}\label{sec:DepositionTechniques}

\paragraph*{\textbf{Thermal evaporation:}} Thermal evaporation is the predominant technique for depositing selenium thin films and has been used in all record-performing devices shown in Fig.~\ref{fig:TimelinePCE}. The high vapor pressure and low melting point of selenium make this approach relatively straightforward. However, as discussed in Section~\ref{sec:StructuralChemistry}, the evaporated molecular beam consists primarily of metastable Se$_\text{8}$ cyclic oligomers. As a result, the as-deposited film is typically amorphous (a-Se), requiring a post-deposition annealing step to form crystalline trigonal selenium (c-Se or t-Se). Cracking these cyclic species before they leave the crucible is challenging, as selenium readily vaporizes before sufficiently high temperatures are reached. Instead, several studies have explored elevated substrate temperatures to promote the dissociation of Se$_8$ rings into more reactive species directly at the surface. Under these conditions, the film has been reported to crystallize during deposition and often exhibits a preferred crystallographic orientation, which will be discussed in Section~\ref{sec:SurfaceWetting}. \\

\paragraph*{\textbf{Electro-deposition:}} To the best of my knowledge, the first report of electro-deposited selenium thin films in photovoltaic devices dates back to 1998, where Tennakone \textit{et al.} deposited selenium into the pores of a nanoporous TiO$_\text{2}$ film by electrolysis of an aqueous selenium dioxide solution \cite{tennakone1998a}. In 2013, Nguyen \textit{et al.} \cite{nguyen2013a} electro-deposited selenium from an aqueous solution containing NaCl, H$_\text{2}$SeO$_\text{3}$, and HCl. Similar to the earlier work, the absorber was deposited onto a mesoporous TiO$_\text{2}$ scaffold, here combined with a compact TiO$_2$ layer inspired by device architectures from dye-sensitized and perovskite solar cells. In this study, multiple deposition pulses with intermediate annealing steps were required to obtain a compact selenium film. The resulting devices reached a power conversion efficiency of 3.0\%.\\

\paragraph*{\textbf{Solution-processing:}} In 2016, Zhu \textit{et al.} reported a fully solution-processed selenium solar cell achieving a PCE of 3.52\% \cite{zhu2016a}. In this work, selenium powder was dissolved in a mixture of ethylenediamine and hydrazine, and the solution was spin-coated onto a mesoscopic TiO$_\text{2}$ scaffold. In 2018, Zhu \textit{et al.} followed up by fabricating the absorber via spin-coating in air \cite{zhu2018a}. While this approach is more practical and potentially more cost-effective, the device efficiency dropped to 1.23\%. In 2022, Deshmukh \textit{et al.} introduced an amine-thiol solvent system to synthesize solution-processed selenium/tellurium alloy films \cite{deshmukh2022a}, which was further explored by Alfieri \textit{et al.} in 2025, demonstrating improved device-level performance. Nevertheless, the 2016 report by Zhu \textit{et al.} remains the most efficient solution-processed selenium solar cell to date.\\

Other techniques for depositing selenium thin films have been demonstrated, though explored less extensively, including closed-space sublimation \cite{shen2025a}, hot-pressing \cite{huang2025a}, glue-bonding of molten selenium \cite{an2025a}, vapor transport \cite{wang2025a, ca2025a}, and blade-coating \cite{lu2022a}.

\begin{figure*}[t!]
    \centering
    \includegraphics[width=\textwidth,trim={0 0 0 0},clip]{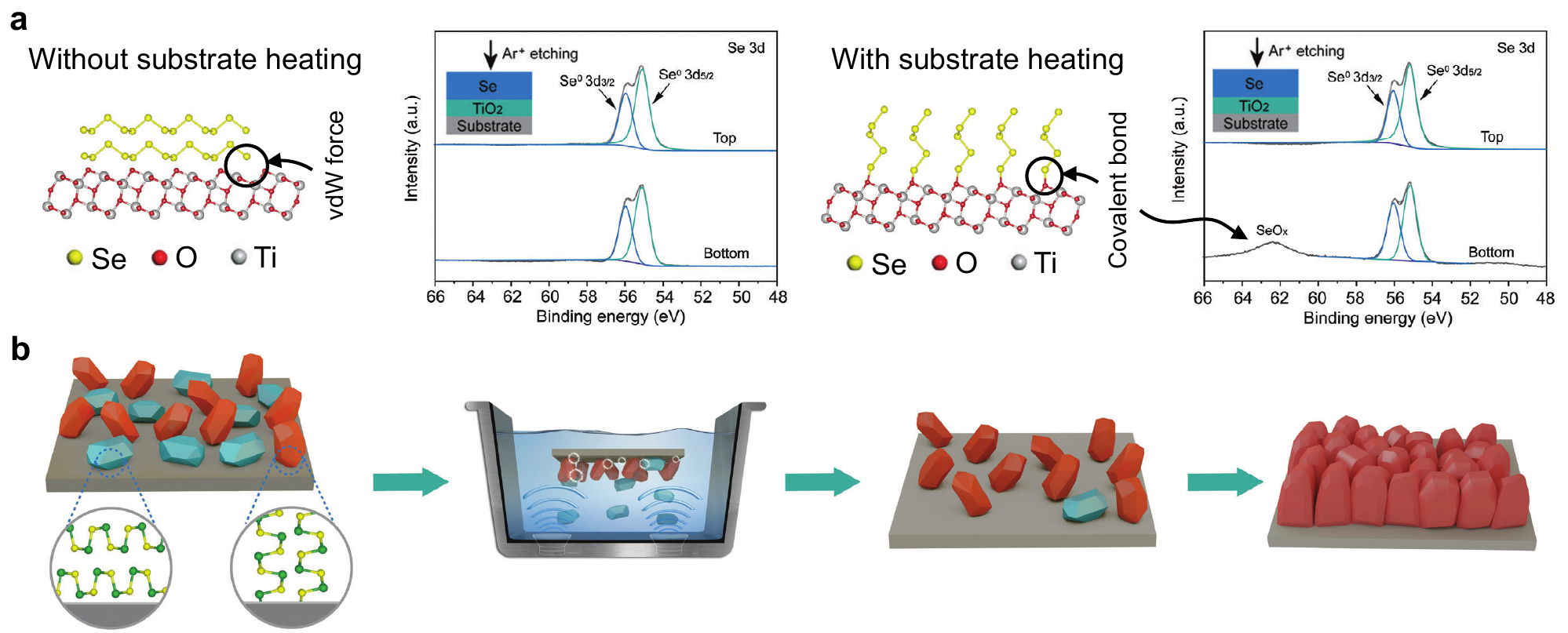}
    \caption{Seed screening strategies for controlling the crystallographic orientation of trigonal selenium thin films. (a) Schematic illustrations of non-preferentially (left) and preferentially (right) aligned trigonal selenium crystals on TiO$_\text{2}$, obtained by thermal evaporation onto room-temperature and preheated substrates, respectively. Corresponding XPS depth profiles of the Se 3d core-level spectra are shown, where a pronounced selenium oxide peak appears at the TiO$_2$ interface only for the preheated sample, indicating covalent bonding through oxygen. Adapted from Ref.~\cite{liu2025a}. (b) Schematic illustration of ultrasonication-assisted seed screening. Adapted from Ref.~\cite{yan2023a}.}
    \label{fig:SeedScreening}
\end{figure*}

\subsection{Surface Wetting}\label{sec:SurfaceWetting}

One of the major challenges in synthesizing selenium thin films is the very low surface energy of trigonal selenium, which promotes dewetting and agglomeration into islands during thermal annealing. The inter-chain surface energy has been calculated to be 0.18~J~m$^\text{-2}$ \cite{kavanagh2025a, wen2026a}, which is in good agreement with experimental values of 0.175~J~m$^\text{-2}$ \cite{lee1971a}. While these surface energetics are intrinsic to the material, several strategies have been developed to mitigate severe dewetting.

In 1984, Hideo \textit{et al.} \cite{ito1984a} deposited selenium thin films on CdSe, which functioned both as an electron-selective contact and as an adhesion layer, enabling the growth of a 500–600~nm thick conformal layer of trigonal selenium. While the authors reported some tendency for the films to peel, careful substrate cleaning largely mitigated this issue, allowing the fabrication of photovoltaic devices with an efficiency of 4.6\% under AM1 illumination.

The dewetting issue was further mitigated by introducing an ultrathin layer of tellurium. To the best of my knowledge, the first report of tellurium in the context of selenium thin films was in 1982 by Kunioka and Nakada \cite{kunioka1982a}, followed by additional works in 1984 \cite{nakada1984a} and 1985 \cite{nakada1985a}, where the $\sim$1~nm tellurium layer was used to “improve adhesion”. While widely adopted in the literature, the precise role of this layer remains unclear. Several factors are worth considering: (1) selenium and tellurium share the same trigonal crystal structure; (2) their similarly low surface energies make selenium less prone to dewetting on tellurium; and (3) the lower vapor pressure and higher melting point of tellurium increase the stability of the ultrathin layer once deposited. Wen \textit{et al.} calculated the surface energies of selenium and commonly used transport layers, along with the corresponding interfacial energies, to predict dewetting behavior \cite{wen2026a}. While selenium is predicted to dewet on all substrates, the interfacial energy with tellurium is significantly lower than with other materials, suggesting that although dewetting may still occur, it is less pronounced when selenium is deposited on tellurium.

In the past decade, a few groups have fabricated conformal selenium thin films without tellurium. Common to these reports is the use of substrate heating, which promotes cracking of the molecular species at the surface and increases the likelihood of covalent bonding between selenium and the substrate. I find this strategy very appealing but still underexplored, as the consequences of removing tellurium from the device architecture have not been fully investigated. Given that tellurium is a narrow-bandgap semiconductor ($E_\mathrm{g}\sim 0.3$~eV), the inclusion of a $\sim$1~nm Te layer at the carrier-separating interface would be expected form a killer interfacial recombination center. Alternatively, tellurium may alloy with selenium to locally reduce the bandgap, as suggested by studies of Se-Te alloys \cite{hadar2019a, deshmukh2022a, alfieri2025a}. While Te is reported to form electrically benign point defects in selenium \cite{kavanagh2025a}, the nature of extended defects, such as clusters, remains unclear. However, when comparing the performance of PV devices fabricated with and without the Te layer, the open-circuit voltage does not improve upon its exclusion. This suggests that the ultrathin interlayer is not contributing significantly to the open-circuit voltage deficit. Further insight could be gained by studying charge density profiles in devices with and without Te using both CV and DLCP, which could clarify how this layer contributes to the relatively high apparent doping density.

\subsection{Controlling Film Texture}\label{sec:FilmTexture}

The anisotropic properties of trigonal selenium motivate growing the absorber film with its helical chains -- and thus covalent bonds -- aligned along the direction of charge transport in photovoltaic devices. A universal strategy widely used to control the crystallographic orientation in other low-dimensional PV materials, such as Sb$_2$Se$_3$, GeSe, and SnSe, is seed screening \cite{zhou2015a, li2019a, otavio2024a}. This approach exploits covalent anchoring: preferentially oriented grains form covalent bonds with the substrate, whereas non-preferentially oriented grains interact only through weaker van der Waals forces and can therefore be removed more easily. In thermal evaporation, these non-preferential seeds are selectively re-evaporated by heating the substrate.

In the case of selenium thin films, seed screening was first explored by Hadar \textit{et al.} in 2019 \cite{hadar2019b}, where selenium was evaporated onto preheated substrates. This produced preferentially oriented grains, but the films exhibited a high density of pinholes and rough morphology. Ultimately, the authors concluded that the overall film quality and morphology are more critical than the crystallographic orientation. More recently, Ding-Jiang Xue’s group refined this approach, showing that precise control of substrate temperature, evaporation rate and surface chemistry is essential for producing more conformal selenium films with preferentially oriented grains \cite{lu2024a, liu2025a, wu2025a}. The formation of covalent bonds between selenium and TiO$_2$ was confirmed via XPS depth profiling, as shown in Fig.~\ref{fig:SeedScreening}(a). The Se 3d core-level spectra reveal a broad selenium oxide peak at the interface with TiO$_\text{2}$ for films deposited on preheated substrates, indicating covalent bonding through oxygen, whereas films deposited at room temperature lack this spectral feature.

In addition to thermal evaporation-based seed screening, Ding-Jiang Xue’s group explored ultrasonication-assisted seed screening to control the orientation  of GeSe and GeS thin films \cite{yan2023a}. Randomly oriented seed crystals are first deposited on the substrate and then placed in an ultrasonic bath, as schematically illustrated in Fig.~\ref{fig:SeedScreening}(b). The ultrasonic stripping force selectively removes the non-preferentially oriented seeds, while those covalently achored to the substrate remain. Notably, the supplementary material (Figure S7 in Ref.~\cite{yan2023a}) demonstrates that this strategy can also be applied to grow preferentially oriented trigonal selenium thin films.

Other notable works include the publication by Shen \textit{et al.} from 2025 \cite{shen2025a}, where selenium thin films were deposited via closed-space sublimation. Interestingly, they found that incorporating small amounts of oxygen during the deposition shifts the optimal substrate temperature for growing preferentially oriented films, facilitating higher-quality growth at lower temperatures. They also identified a sweet spot in the oxygen partial pressure for achieving compact films with a desirable morphology. Earlier studies by Xerox Corporation explored the effect of both chlorine and oxygen on the nucleation and crystallization of amorphous selenium thin films in the context of xerographic applications \cite{jansen1980a}. In 2026, Xia \textit{et al.} \cite{xia2026a} demonstrated that increasing the vapor pressure of selenium during the thermal annealing step can suppress sublimation, promoting re-evaporation and re-deposition, and ultimately resulting in preferentially oriented films. Finally, in the publication by Cano \textit{et al.} from 2025 \cite{ca2025a} preferentially oriented selenium films were grown via vapor transport deposition. This work stands out for its comprehensive analysis of crystallographic orientation using both texture coefficients, integrated reflection intensity ratios, and pole figure analysis.

\begin{figure*}[t!]
    \centering
    \includegraphics[width=\textwidth,trim={0 0 0 0},clip]{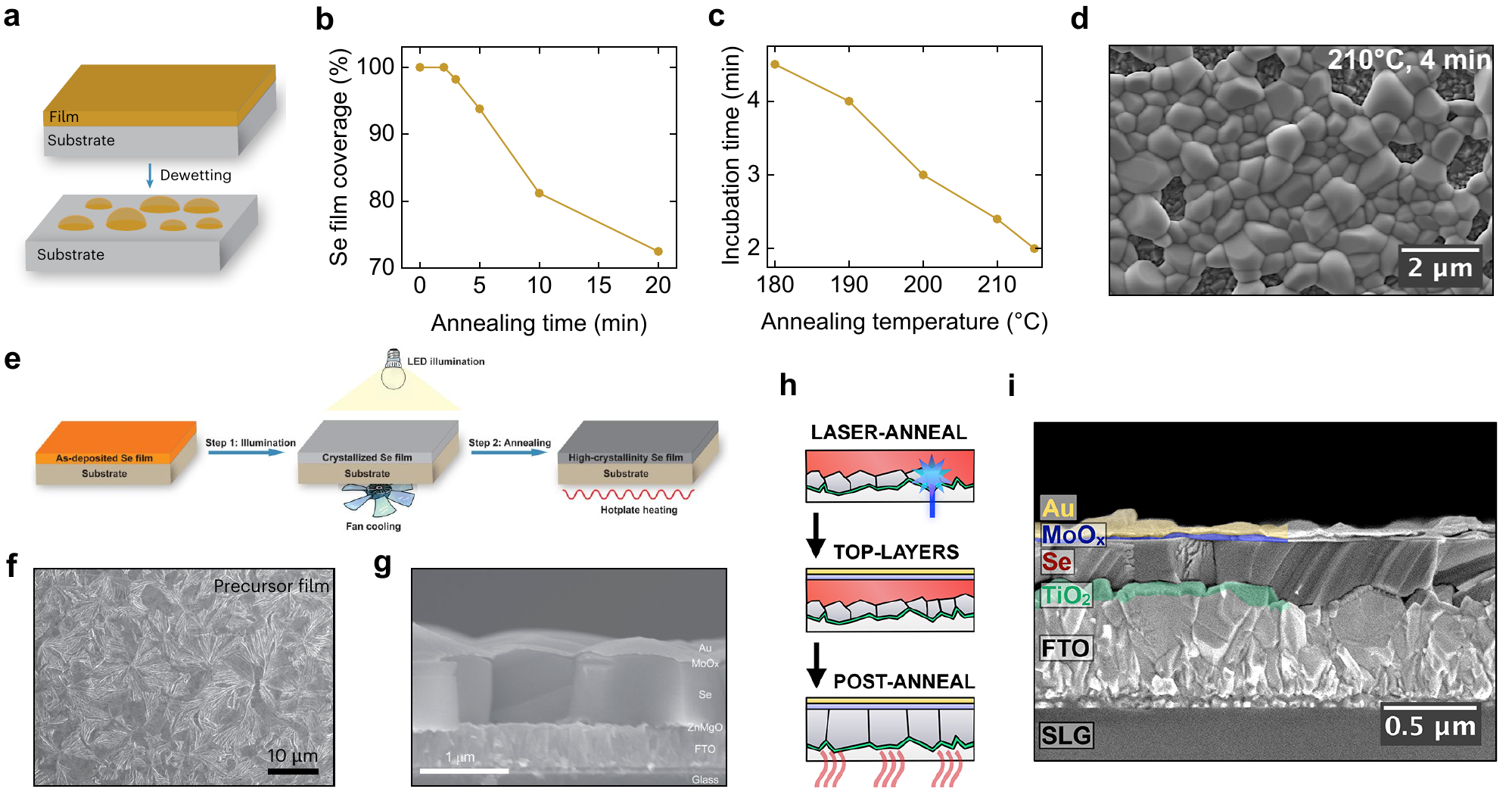}
    \caption{Strategies for crystallizing selenium thin films.  (a) Schematic illustration of dewetting, where a continuous as-deposited selenium film agglomerates into isolated islands. (b) Selenium film coverage on a ZnMgO substrate as a function of annealing time at 200~C. (c) Incubation time for dewetting of selenium as a function of annealing temperature. (a-c) Adapted from Ref.~\cite{wen2026a}. (d) Top-view SEM image of a selenium thin film annealed at 210~C for 4~min from Ref.~\cite{nielsen2024b}, exceeding the incubation time which resulted in the formation of pinholes. (e) Schematic illustration of the illumination-assisted annealing (IAA) strategy. (f) Top-view SEM image of the precursor film immediately after the 12-hour photocrystallization step. (g) Cross-sectional SEM image of the IAA-processed device after a final thermal annealing step. (e-g) Adapted from Ref.~\cite{wen2026a}. (h) Schematic illustration of the laser-annealing strategy. (i) Cross-sectional SEM image of the laser-annealed device after a final post-annealing step. (h-i) Adapted from Ref.~\cite{nielsen2023a}.}
    \label{fig:CrystallizationStrategies}
\end{figure*}

\subsection{Crystallization Strategies}

As discussed in Section~\ref{sec:DepositionTechniques}, the predominant approach for synthesizing selenium thin films is a two-step process, where an amorphous selenium layer is first deposited by thermal evaporation, followed by a crystallization step. Even in cases where the as-deposited film is already crystalline, such as electro-deposition, a subsequent processing step is typically still carried out to further improve the crystallinity of the absorber layer.\\

\paragraph*{\textbf{Thermal annealing:}} The conventional strategy for crystallizing amorphous selenium thin films is thermal annealing on a pre-heated hotplate in air. While this process may appear straightforward, it involves several competing mechanisms. First, the high vapor pressure of selenium leads to significant sublimation even at atmospheric pressures \cite{nielsen2024b}, and since selenium oxide exhibit an even higher vapor pressure, an oxidative environment can further increase potential material loss \cite{nielsen2025b}. Second, as discussed in Section~\ref{sec:SurfaceWetting}, the low surface energy promotes dewetting, which has been schematically illustrated in Fig.~\ref{fig:CrystallizationStrategies}(a). Achieving a conformal, high-quality thin film with large grains therefore requires a delicate balance between annealing temperature and time, where even small variations on the order of a single degree or a few seconds can have a significant impact. This interplay was quantified by Wen \textit{et al.} \cite{wen2026a}, who studied selenium film coverage as a function of annealing time and incubation time as a function of temperature, as shown in Fig.~\ref{fig:CrystalStructures}(b) and (c), respectively. Here, the incubation time is defined as the onset of pinhole formation. These results very nicely explain why the optimal annealing conditions are typically reported around 190~C for 4~min or 200~C for 2~min. Increasing either the temperature or annealing time leads to the formation of large grains, but at the expense of pinhole formation, as demonstrated in the top-view SEM image in Fig.~\ref{fig:CrystallizationStrategies}(d).

Beyond annealing time and temperature, the processing parameter space has been explored only to a limited extent. Even the methodology used to define the annealing temperature is often unclear, i.e. whether it refers to the actual surface temperature, the hotplate setpoint, or the temperature measured by an internal thermocouple. At the same time, there are clear indications that additional parameters play an important, yet poorly understood, role. For example, studies by Jansen \cite{jansen1980a} and Shen \textit{et al.} \cite{shen2025a} show that oxygen influences the crystallization kinetics of selenium, although the underlying mechanism remains unclear. Furthermore, sublimation during annealing directly impacts film conformality through pinhole formation, but further studies limited to annealing time and temperature alone are unlikely to provide substantial new insight. Instead, a more systematic exploration of the processing environment -- such as controlled gas compositions including for example only inert gases or selenium vapors, or elevated pressures above 1~bar -- could help clarify the role of oxygen, water, and other reactive species, while also mitigating selenium loss during crystallization.\\

\paragraph*{\textbf{Illumination-assisted annealing:}} To suppress dewetting while promoting the growth of large crystal grains in a conformal selenium thin film, Wen \textit{et al.} introduced an illumination-assisted annealing (IAA) strategy. This method produced very large grains of $\sim2.7$~$\upmu$m and a record device efficiency of 10.3\%. As schematically illustrated in Fig.~\ref{fig:CrystallizationStrategies}(e), the approach involves illuminating the as-deposited amorphous film with an LED filtered to remove infrared light, which minimizes sample heating. To further ensure that the film remains at room temperature, a fan was used during the 12-hour photocrystallization step, effectively suppressing dewetting. In this process, it is the photogenerated carriers in the amorphous selenium film that drive the growth of the crystallites. The resulting precursor film exhibits a rather peculiar structure and morphology, as seen in the top-view SEM image in Fig.~\ref{fig:CrystallizationStrategies}(f). One detail not captured in the schematic is that a layer of MoO$_\text{x}$ and Au is deposited prior to a final thermal annealing step on a hotplate, which further improves crystallization. This post-processing step, referred to here as CSA, is discussed in Section~\ref{sec:PostProcessing}. The SEM cross-section of the complete photovoltaic device in Fig.~\ref{fig:CrystallizationStrategies}(g) shows that the large grains extend through the full 1~$\upmu$m thickness of the absorber, while the surface roughness remains relatively low.

The use of light to assist the crystallization of selenium thin films in photovoltaic devices was, to the best of my knowledge, first demonstrated by Hadar \textit{et al.} in 2019 \cite{hadar2019b}. In their work, illuminating the amorphous film during thermal annealing reduced the crystallization temperature by $\approx$20~C and the optimal annealing temperature by $\approx$5~C. In the context of the incubation time as a function of annealing temperature, shown in Fig.~\ref{fig:CrystallizationStrategies}(c), this allowed for longer annealing time without detrimental pinhole formation, producing larger grains and improving device efficiencies. These results suggest that photogenerated carriers significantly enhance the crystallization kinetics. Earlier fundamental studies provide insight into this effect. In 1968, Dresner and Stringfellow \cite{dresner1968a} demonstrated that photocrystallization of amorphous selenium is driven by the generation of electron-hole pairs rather than the absorbed optical or thermal energy. Studying the topography of the energy bands, they found that small crystallites in the amorphous film act as sinks for holes, and the crystal growth rate is governed by the flux of free holes to the amorphous-crystalline interface. In 1974, Clement \textit{et al.} \cite{clement1974a} complemented this work by investigating the role of the substrate, comparing selenium films photocrystallized on glass and gold. They observed crystallization inhibition near the gold interface, which they attributed to a space-charge region that repels holes, highlighting the dependence of nucleation on the carrier density near the substrate.\\

\paragraph*{\textbf{Laser-crystallization:}} Another approach to crystallize selenium thin films using light is laser annealing \cite{nielsen2023a}. In this method, a buried crystalline seed layer is first created by photo-exciting the as-deposited amorphous selenium through a semi-transparent substrate. The subsequent layers of the photovoltaic device are then deposited, after which the remaining amorphous selenium is crystallized from the buried seed layer using a thermal annealing step. This is schematically illustrated in Fig.~\ref{fig:CrystallizationStrategies}(h). Since the film grows from crystalline seeds in intimate contact with the surrounding amorphous phase, this process corresponds to solid-phase epitaxy.The resulting device, shown in the cross-sectional SEM-image in Fig.~\ref{fig:CrystallizationStrategies}(i), exhibits a remarkably flat surface morphology, enabling high carrier-collection efficiencies and a then-record fill factor of 63.7\%. The original motivation for this approach was twofold. First, laser-induced crystallization was proposed as a lever to shift the electron chemical potential without introducing extrinsic dopants, thereby altering the formation enthalpy of both intrinsic and extrinsic defects \cite{kim2020a}, although this effect was not investigated experimentally. Second, we wanted to create a moving molten zone, analogous to a floating-zone process, and “drag” a single crystal through the film in the wake of the laser. While this was not achieved in the initial demonstrations, it is still an interesting direction for future work.\\

Overall, light-assisted crystallization -- whether through flood illumination or a localized laser beam -- has emerged as a promising strategy for producing high-quality crystalline selenium thin films. Compared to conventional thermal annealing, where the optimal processing window is extremely narrow due to short incubation times at elevated temperatures, photo-induced crystallization offers greater control by inducing energy very accurately and localized in both time and space. This enables crystallization at lower effective temperatures without heating the substrate, potentially broadening the range of compatible substrate materials and improving process scalability for mass production of selenium solar cells with low cost. However, in all reported cases, a subsequent thermal post-processing step is still used to further enhance crystallinity.

\subsection{Post-Processing}\label{sec:PostProcessing}

Many device-level studies on selenium solar cells include a thermal post-processing step after completing the full device stack. In a work of ours from 2024 \cite{nielsen2024b}, we demonstrated that the encapsulating nature of the deposited top-layers suppresses the sublimation of selenium, enabling thermal annealing at higher temperatures and for longer durations than when selenium is exposed to air. We termed this approach "closed-space annealing" (CSA) and showed that it significantly improves carrier-collection efficiency. To our surprise, the enhanced collection of longer-wavelength photons observed after CSA could not be attributed to an increased diffusion length in the absorber. Instead, we hypothesized that the absorber undergoes partial recrystallization, allowing adjacent grains to merge while previously delaminated regions -- possibly formed due to the high surface roughness and low surface energy of trigonal selenium -- are given a second change to re-establish proper electrical contact with the absorber. Such a process would effectively reduces the charge-transfer resistance, consistent with impedance spectroscopy measurements.

Two additional publications relevant to thermal annealing of the complete device stack are those by Bao \textit{et al.} \cite{bao2025a} and Segura-Blanch \textit{et al.} \cite{segura-blanch2025a}. In these two independent works, the interfacial chemistry between selenium and MoO$_\text{x}$ was investigated, and the formation of a MoSe$\text{x}$ interlayer was reported. This suggests that the post-processing thermal annealing step may promote the formation of a selenized interface, which has been proposed to facilitate hole extraction and suppress interfacial recombination.

In an independent study published around the same time as the introduction of CSA, Lu \textit{et al.} proposed a critical melting-annealing (CMA) strategy, in which the annealing temperature is pushed close to the melting point of selenium to supply sufficient energy to overcome the activation barrier for incorporating disordered selenium chains into the lattice \cite{lu2024a}. However, it is not clear to me how thermal annealing at temperatures as high as 215~C was achieved without significant sublimation losses. According to their reported methodology, this CMA step was performed after deposition of the Au contacts in complete devices, suggesting that encapsulation may play a role. But this does not explain how the cross-sectional SEM images presented in the supplementary material show conformal films for both control and CMA-treated samples without encapsulating layers. While these results are very impressive, the reported processing conditions and resulting film morphology appear inconsistent with both prior reports from other groups, as well as with their own subsequent study from 2026 by Wen \textit{et al.} \cite{wen2026a}, where dewetting, incubation time, and sublimation were systematically investigated.

\begin{figure*}[t!]
    \centering
    \includegraphics[width=\textwidth,trim={0 0 0 0},clip]{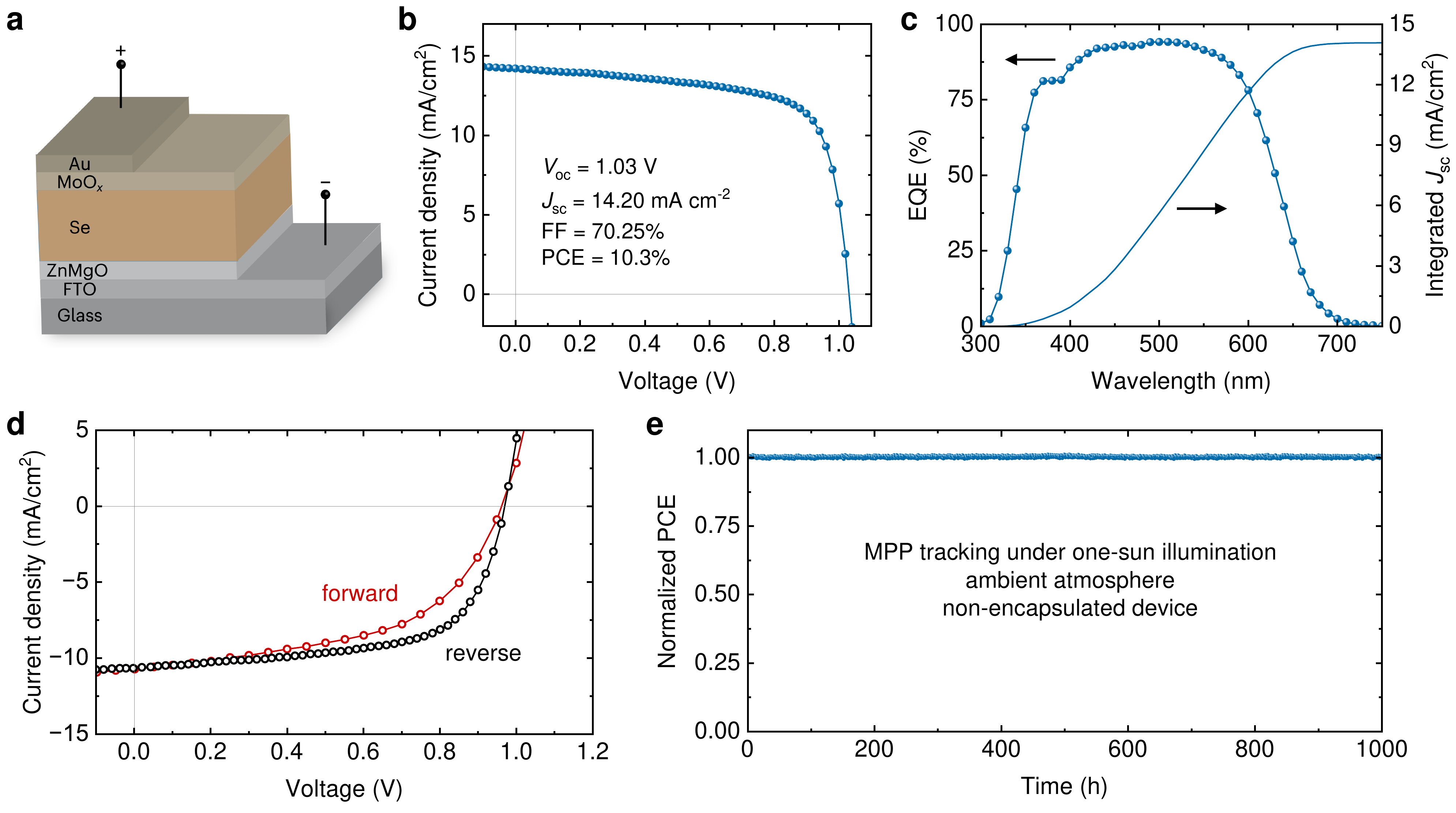}
    \caption{Photovoltaic device performance of selenium thin film solar cells. (a) Schematic illustration of the device architecture of the current record device. (b) Certified J–V curve of the current record device. (c) Certified EQE spectrum and the corresponding integrated short-circuit current density as a function of wavelength. (a–c) Adapted from Ref.~\cite{wen2026a}. (d) Forward- and reverse-scanned J-V curves of the previous record device from Todorov \textit{et al.}~\cite{todorov2017a}, featuring the same device architecture as shown in (a). (e) Normalized power conversion efficiency under maximum power point tracking at one-sun illumination as a function of time for the current record device. The measurement was performed in ambient atmosphere without encapsulation. Adapted from Ref.~\cite{wen2026a}.}
    \label{fig:PVDevices}
\end{figure*}

\section{Photovoltaic Devices}

The review papers from 2019 by Zhu \textit{et al.} \cite{zhu2019a} and 2026 by Shang \textit{et al.} \cite{shang2026a} very nicely summarize the device architectures explored for selenium solar cells. Despite significant improvements in reported efficiencies since 2019, the best-performing devices still rely on the structure first demonstrated by Todorov \textit{et al.} in 2017 \cite{todorov2017a}. This configuration, FTO/ZnMgO/Te/Se/MoO$_\text{x}$/Au, is schematically illustrated in Fig.~\ref{fig:PVDevices}(a). Notably, recent efficiency records are now certified, reflecting the increasing maturity of the selenium PV field. The certified J–V curves (forward and reverse scans) and EQE spectrum of the 10.3\% device reported by Wei \textit{et al.} are shown in Fig.~\ref{fig:PVDevices}(b) and (c), respectively.

The most widely used device structure, however, is still the standard architecture introduced by Nakada and Kunioka \cite{nakada1985a}, which uses TiO$\text{2}$ as the electron-selective contact. Devices based on this configuration typically demonstrate open-circuit voltages up to $\sim$100~mV lower than those using ZnMgO. This trend is consistent with the band alignment reported by Todorov \textit{et al.} \cite{todorov2017a}, which is shown in Fig.~\ref{fig:BandAlignmentFigure}(c), indicating a larger conduction band offset at the TiO$_\text{2}$/Se-interface. However, these results are inconsistent with the ZnMgO/Se band alignment measurements from our group reported in Ref.~\cite{nielsen2024a}. The validity and associated uncertainties of these measurements were discussed in details in Section~\ref{sec:BandPositions}.

Based on SCAPS-1D simulations and the critical assessment of the material properties, I personally consider the carrier lifetime to be the main cause of the pronounced open-circuit voltage loss in selenium devices. With certified short-circuit current densities approaching the SQ limit and fill factors exceeding 70\%, I would not anticipate significant improvements from further device-level engineering. Instead, the priority should be to enhance the PL quantum yield and the carrier lifetime through absorber engineering. Guided by this perspective, I have kept this section relatively concise, highlighting only a few select studies on carrier-selective contacts, JV-curve hysteresis, and operational stability, while referring readers to the reviews by Zhu \textit{et al.} and Shang \textit{et al.} for a more comprehensive discussion of device-level approaches. Finally, as an outlook, I briefly note other potential applications of selenium thin films and solar cells that I consider relevant to the PV community.

\subsection{Carrier-Selective Contact Materials}

The study by Todorov \textit{et al.} from 2017 \cite{todorov2017a} mainly focused on improving the device architecture introduced by Nakada and Kunioka in 1985 \cite{nakada1985a}. The original structure, FTO/TiO$_\text{2}$/Te/Se/Au, lacked a hole-selective contact, leading to the formation of a Schottky barrier at the back contact that acted as a major site of interfacial recombination. To address this, the authors introduced the high work-function transition metal oxide MoO$\text{x}$ as a hole-selective layer. The high work-function of MoO$_\text{x}$ sets up a back-surface field in the absorber that repels electrons, while the substoichiometry of MoO$_\text{x}$ creates an oxygen-vacancy defect band that facilitates hole transport \cite{battaglia2014a}. As the carrier transport is not driven by quantum tunneling, an optimal thickness of $\approx$15 nm was found, likely balancing charge transport resistance and interface passivation quality. This modification reduced carrier recombination at the back contact and improved overall carrier collection. As discussed in Section~\ref{sec:PostProcessing}, Bao \textit{et al.} \cite{bao2025a} and Segura-Blanch \textit{et al.} \cite{segura-blanch2025a} independently studied the materials chemistry at the Se/MoO$_\text{x}$-interface, observing the formation of a MoSe$_\text{x}$ interlayer. Segura-Blanch \textit{et al.} also evaluated other high work-function oxides, including V$_\text{2}$O$_\text{x}$ and WO$_\text{x}$, finding MoO$_\text{x}$ and V$_\text{2}$O$_\text{x}$ to be the most effective HTL candidates, achieving device efficiencies of 5.5\% and 5.3\%, respectively, under AM1.5G illumination.

Inspired by advances from the field of chalcogenide photovoltaics \cite{hariskos2009a}, Todorov \textit{et al.} also introduced a tunable bandgap buffer layer, Zn$_\text{x}$Mg$_\text{1-x}$O, as a replacement for TiO$_\text{2}$. This substitution improved the conduction band alignment with selenium, leading to a significant increase in open-circuit voltage. However, the high resistivity of MgO makes this material a double-edged sword: while higher Mg content further increased the open-circuit voltage, it simultaneously reduced carrier collection and overall device efficiency. As a compromise, the optimal composition was identified to be approximately Mg/(Zn+Mg) = 0.1 with a thickness in the range of 60-85 nm.

Another approach could be to include ultra-thin molecular or polymeric dipole layers. While the conduction band edges of TiO$_\text{2}$ or ZnMgO may not be ideally aligned with selenium, an interfacial dipole layer can modify the local band alignment by up to several electron volts. This strategy was suggested by prof. A. Walsh in the concluding remarks of a Faraday Discussion on emerging photovoltaics \cite{walsh2022a}, as an alternative to directly alloying buffer layers and living with the trade-offs between bandgap, band alignment, and resistivity.

\subsection{Hysteresis}

Hysteresis in JV-curve measurements is particularly well-known from the field of perovskite solar cells and is often linked to ion migration, charge trapping, or ferroelectric polarization. This non-ideal behavior can sometimes be mitigated through alternate device architectures or improved interface quality. In the case of selenium solar cells, the first report showing both forward and reverse scans is, to the best of my knowledge, the work by Todorov \textit{et al.} from 2017 \cite{todorov2017a}, where hysteresis was observed. These scan-dependent JV-curves are shown in Fig.~\ref{fig:PVDevices}(d). In 2019, Hadar \textit{et al.} also reported scan-dependent JV-curves, with higher open-circuit voltages during reverse sweeps. With this publication, hysteresis have been documented for selenium solar cells using both TiO$_\text{2}$ and ZnMgO as the electron-selective contact. Several of the later studies from Ding-Jiang Xue’s group use nominally identical device architectures, but demonstrate no observable hysteresis. This includes the certified record device, shown in Fig.~\ref{fig:PVDevices}(b), suggesting that JV-curve hysteresis is not intrinsic to selenium solar cells or the champion device architecture, and that the previously observed scan-dependence likely arises from non-idealities at interfaces. Nevertheless, reporting both scan directions remains an important practice, as hysteresis has been documented in several independent studies for selenium solar cells with efficiencies exceeding 5\%.

\begin{figure*}[t!]
    \centering
    \includegraphics[width=\textwidth,trim={0 0 0 0},clip]{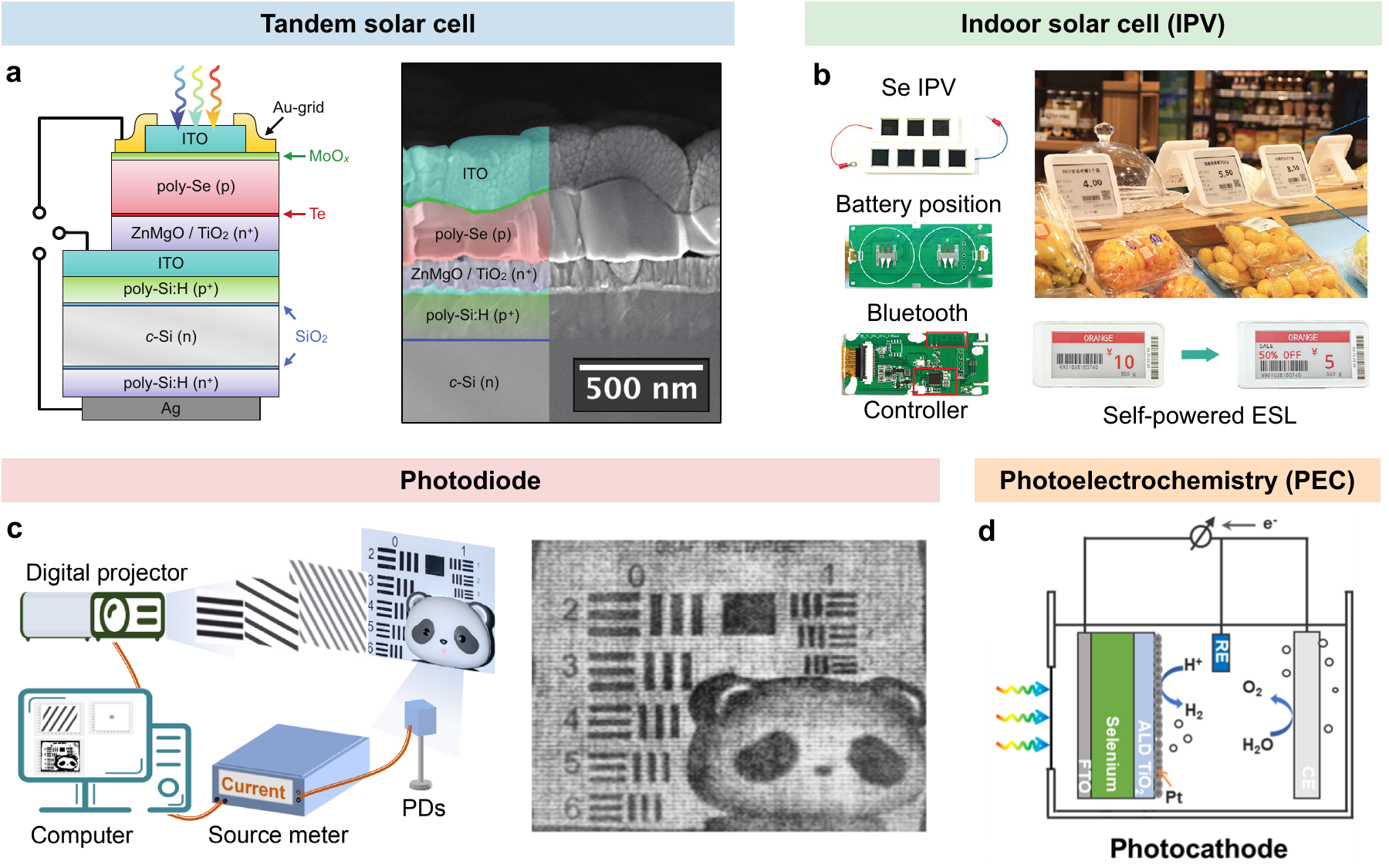}
    \caption{Derivative applications of selenium thin films beyond single-junction PV. (a) Schematic and cross-sectional SEM image of a monolithic selenium/silicon tandem solar cell, adapted from Ref.~\cite{nielsen2024a}. (b) Mini-module of selenium indoor solar cells integrated into self-powered wireless electronic shelf labels for supermarket applications, adapted from Ref.~\cite{lu2024a}. (c) Selenium-based photodiode integrated into a single-pixel imaging system used for image reconstruction, from Ref.~\cite{bai2024a}. (d) Selenium-based photoelectrochemical (PEC) water-splitting device featuring an ALD-deposited TiO$_\text{2}$ layer, adapted from Ref.~\cite{li2021a}.}
    \label{fig:PVApplications}
\end{figure*}

\subsection{Operational Stability}

While power conversion efficiency is typically the spotlight metric, stability is ultimately more important for assessing technological readiness and commercialization potential. Trigonal selenium thin films are generally cited as air- and moisture-stable, with references tracing back to Zhu \textit{et al.} \cite{zhu2016a} and Liu \textit{et al.} \cite{liu2020a}. However, only more recent reports provide actual operational stability data \cite{nielsen2024b, xia2026a, wen2026a, duan2026a, liu2025a, an2025a, lu2024a, wei2023a, lu2022a, yan2022a, liu2020a}. It is encouraging that these studies consistently report non-encapsulated devices measured in an ambient atmosphere, with negligible observable degradation in most cases. This includes the 10.3\% certified record device from Wen \textit{et al.} \cite{wen2026a}, shown in Fig.~\ref{fig:PVDevices}(e). Two points of critique: first, the cell temperature is not reported to be tracked, which could easily have been done by mounting a thermocouple on a dummy device in parallel to the device under test. Second, the reported PCEs are most often normalized for long-term measurements, which is a common practice I find unjustifiable unless you are deliberately trying to hide something. If normalization is considered essential, a secondary y-axis showing the absolute efficiency should be included; otherwise, it is only reasonable to assume that the device either degraded prior to the stability measurement or that a parallel-processed device with lower absolute efficiency was measured, showing no apparent degradation.

As a final comment on measuring and reporting operational stability, I think we should adopt the testing protocols from the field of perovskite photovoltaics, where stability is widely recognized as the main bottleneck to commercialization. To properly assess environmental and operational stability, testing conditions should exceed expected real-world exposure. This means increasing humidity and temperature until the device reaches its breaking point, diagnosing the mechanisms behind degradation, and then engineering the absorber and device to withstand even harsher conditions. By challenging operational stability at an early stage in research, selenium solar cells can be positioned as an even more serious candidate for commercial applications.

\subsection{Applications}

Throughout this review, I have focused exclusively on selenium thin films in the context of single-junction photovoltaic devices. However, many aspects of device structures and thin-film processing are shared with other derivative applications, making it useful to consider what other groups are doing in related areas, such as photodetectors, thermoelectrics, and photoelectrochemistry. In this subsection, I briefly summarize a selection of publications from these application spaces that I believe provide valuable insights into selenium thin films, which are transferable to both fundamental studies and applied research in selenium solar cells.\\

\paragraph*{\textbf{Bifacial solar cells:}} To date, only a few studies have explored bifacial selenium solar cells \cite{youngman2021a, nielsen2024a, an2025a}. Among these, the work by Youngman \textit{et al.} is particularly insightful, as the carrier collection efficiency under illumination from the opposite side of the junction, studied as a function of absorber thickness, provides direct insight into the low carrier diffusion lengths in selenium. In addition, replacing the metallic back contact with a transparent conductive oxide presents challenges that are directly relevant for tandem integration, where both contact materials must be transparent. In this context, the work by An \textit{et al.} offers a promising approach through a glue-bonding strategy, demonstrating highly efficient devices with a bifaciality factor of 90.1\% \cite{an2025a}.\\

\paragraph*{\textbf{Tandem solar cells:}} Our group's work from 2024 remains, to date, the only experimental demonstration of a selenium-based tandem solar cell. The selenium/silicon monolithic tandem device is shown schematically in Fig.~\ref{fig:PVApplications}(a) alongside a cross-sectional SEM image. In this study, we motivated an industrially relevant application of selenium photovoltaics and demonstrated how SCAPS-1D simulations of band alignment could guide a tenfold improvement in device efficiency. At the same time, several key challenges were identified, including the need for polarity inversion and the implementation of contact grids on transparent ohmic contacts. To date, polarity inversion in photovoltaic devices has only been demonstrated by Liu \textit{et al.} \cite{liu2020b} using organic transport layers (PEDOT:PSS and PCBM), and by An \textit{et al.} through glue-bonding of molten selenium \cite{an2025a}. Neither of these approaches has yet been implemented in a monolithic tandem architecture.\\

\paragraph*{\textbf{Indoor solar cells:}} With the emergence of the internet-of-things (IoT), indoor photovoltaics have attracted a lot of attention. Given that the optical bandgap of trigonal selenium is well matched to the emission spectrum of white LEDs, selenium solar cells are particularly suitable for indoor energy harvesting. As a result, several groups have explored selenium-based indoor solar cells, with device architectures no different than those used for outdoor applications \cite{bishop2017a, yan2022a, wang2024a, wei2023a}. A representative field demonstration from Lu \textit{et al.} is shown in Fig.~\ref{fig:PVApplications}(b), where a mini-module of selenium devices is integrated into self-powered wireless electronic shelf labels used in supermarkets \cite{lu2024a}.\\

\paragraph*{\textbf{Photodiodes:}} There are also quite a few papers exploring selenium-based photodiodes \cite{li2024a, chen2024a, adachi2023a, qin2017a, hu2016a, zheng2017a, chang2019a}. An example is shown in Fig.~\ref{fig:PVApplications}(c), adapted from Bai \textit{et al.} \cite{bai2024a}, where a selenium-based single-pixel imaging system is used to reconstruct a digitally projected image. Another particularly insightful study by Chen \textit{et al.} investigated deep trap levels in selenium and SeTe thin films using deep-level transient spectroscopy (DLTS). Interestingly, the fitted DLTS signal amplitude exhibits a linear dependence on the filling pulse width, suggesting that the dominant trap states originate from extended defects rather than point-like defects. This observation aligns with the hypothesis proposed by Kavanagh \textit{et al.} \cite{kavanagh2025a}, which suggests that while trigonal selenium is tolerant to point defects, its structurally flexible nature makes it susceptible to the formation of extended defects that may limit photovoltaic performance.\\

\paragraph*{\textbf{Photoelectrochemistry:}} A number of studies have explored selenium as an absorber material in photoelectrochemical systems \cite{gissler1980a, qian2012a, li2021a}. In this context, environmental stability is pushed to much harsher limits, which is worth noting from a photovoltaic perspective. Of particular interest is the water-splitting device shown in Fig.~\ref{fig:PVDevices}(d), where atomic layer deposited (ALD) TiO$_\text{2}$ serves both as a protection layer and as an electron-selective contact. While TiO$_\text{2}$ itself is not new in selenium photovoltaics, this architecture demonstrates that ALD-deposited TiO$_\text{2}$ can also be used to achieve polarity inversion.\\

\paragraph*{\textbf{Thermoelectrics:}} Due to its low-dimensional crystal structure and charge transport properties, trigonal selenium is also being explored for thermoelectric applications \cite{zhang2021a, peng2021a, ramirez-montes2024a}. This field offers valuable insights for the photovoltaic community, particularly in controlling crystallographic orientation and understanding how the underlying structural chemistry influences carrier dynamics, including thermal conductivity, anisotropy, and electron-phonon interactions.

\section{Challenges and Open Questions}

It has been an absolute pleasure to work through the literature on selenium thin films and solar cells, and to see how the field has developed -- not only over the past centuries, but especially over the past decade. While significant progress has been made in both our fundamental understanding of the structural chemistry and material properties, as well as in device engineering and demonstrated photovoltaic performance, a range of challenges and open questions still remain. Addressing these will be essential for further improving the efficiency and PV performance potential of selenium solar cells. Although many of these points have been discussed throughout this review, this section summarizes the key challenges and open questions highlighted along the way.

\begin{figure*}[t!]
    \centering
    \includegraphics[width=\textwidth,trim={0 0 0 0},clip]{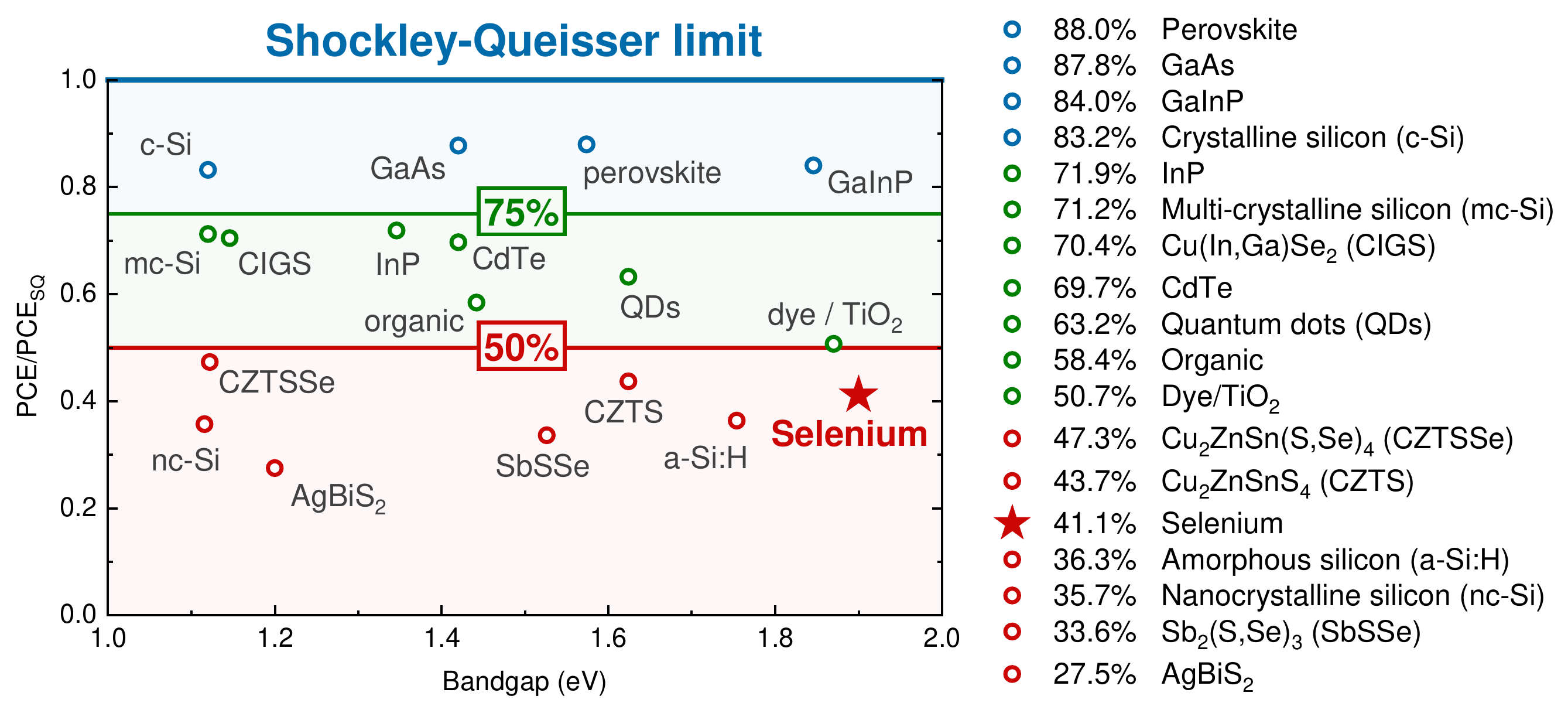}
    \caption{The power conversion efficiency of various photovoltaic materials relative to their Shockley–Queisser (SQ) limit as a function of bandgap. The data is sourced from Ref.~\cite{green2025a}.}
    \label{fig:PVRelativeToSQLimit}
\end{figure*}

\begin{itemize}
    \item While substrate heating has been explored as a means to crack monoclinic selenium rings upon arrival at the substrate surface, pre-cracking the molecular beam before low-reactivity selenium species reach the substrate remains unexplored in the context of selenium thin-film growth. In particular, it would be valuable to investigate the composition of the molecular beam as a function of process conditions.
    \item It remains unclear whether individual trigonal selenium grains consist of helices with a single handedness or a mixture of left- and right-handed screw directions.
    \item The static dielectric constant has yet to be properly determined using capacitance-based measurements, which would provide access to lower-frequency responses than standard optical techniques.
    \item While recent advances have enabled PL spectroscopy on selenium thin films, the Urbach energy remains unexplored. Resolving the Urbach tail into its dynamic and static components would provide insight into how structural disorder may intrinsically limit the attainable optoelectronic quality.
    \item Despite the very low signal-to-noise ratios observed in steady-state PL measurements, the PL quantum yield has yet to be quantified. Systematic studies across different carrier-selective contacts and absorber thicknesses could provide valuable insight into the role of surface and interface defects, and help assess the need for targeted passivation strategies to further improve device performance.
    \item The hypothesis of carrier localization on the picosecond timescale into long-lived trap states has yet to be conclusively verified. This could be tested by designing a THz pump–probe–push experiment, where a lower-energy push pulse promotes localized carriers back into delocalized states, transiently restoring high carrier mobilities.
    \item Several studies report a characteristic carrier dynamic on the picosecond timescale, and some also note the presence of a relatively deep acceptor level. However, the nature and origin of this acceptor remains unresolved.   
    \item If the doping density is indeed on the order of $p \sim 10^{16}$~cm$^{-3}$, the origin of this hole concentration remains unexplained. Initial DFT studies, could not reproduce such carrier densities by introducing high concentrations of extrinsic point defects.
    \item We do not yet understand the full impact of the ultrathin tellurium adhesion layer on device performance. Comparisons of parallel-processed devices with and without this layer of tellurium show negligible change in open-circuit voltage, suggesting it does not form a killer interfacial recombination center. However, its effect on apparent doping and interfacial carrier dynamics remains uncertain and could be clarified through complementary CV and DLCP measurements.
    \item The main loss in selenium solar cells remains the severe open-circuit voltage deficit. While low effective carrier lifetimes likely account for most of this loss, some groups continue to point to non-ideal carrier-selective transport layers. While I do not share this opinion, it would be valuable to examine the interfacial energy band alignments more carefully -- particularly at the ZnMgO/Se-interface across samples from different laboratories.
    \item While ultrashort free carrier lifetimes and high doping densities remain a working hypothesis based on multiple independent studies, it has yet to be shown that an alternative set of experimentally justified material parameters can allow drift-diffusion simulations to reproduce experimental JV-curves and EQE spectra. This is particularly relevant for lower doping densities, which remain controversial. 
    \item DLTS measurements and DFT studies collectively suggest that extended defects may be problematic, but this has yet to be investigated in detail. A deeper understanding is needed before mitigation strategies can be designed to target specific types of structural defects.
    \item Preferentially oriented growth of selenium thin films without a tellurium layer has so far only been demonstrated on TiO$_\text{2}$ substrates. Achieving similar oriented growth on other substrates, including ZnMgO used in the current record-efficiency device, remains an open challenge.
    \item Laser annealing has proven to be an effective approach for crystallizing selenium thin films. However, it has yet to be demonstrated whether a single crystal selenium thin film can be drawn from the wake of the laser through a floating zone of selenium.
    \item Polarity inversion, i.e., fabricating the electron-selective contact on top of the selenium thin film, is still underexplored. Organic transport layers offer promising routes for this, but they have yet to be implemented in for example selenium-based tandem solar cells, which is expected to significantly enhance the overall device efficiency.
\end{itemize}

\section{Concluding Remarks}

The 2026 Nature Energy publication from Wen \textit{et al.} \cite{wen2026a} marked the first time selenium solar cells broke through the 10\% efficiency barrier -- certified! This is impressive in its own right, but even more so considering the wide bandgap of selenium, which imposes a lower detailed-balance efficiency limit compared to lower-bandgap absorbers like crystalline silicon or GaAs. Fig.~\ref{fig:PVRelativeToSQLimit} shows the PCE of various PV materials relative to their bandgap-dependent Shockley–Queisser limit, illustrating that selenium PV is already on par with kesterites and outperforms amorphous silicon, despite decades of research and industrial development.

I find these relatively high efficiencies remarkable, particularly given the difficulty of measuring PL on selenium thin films. To me, this suggests that the most promising path to further improve device performance lies in enhancing the optoelectronic quality of the absorber layer. The FOM by Crovetto reinforces this, showing that the device quality factor currently exceeds the absorber quality factor \cite{crovetto2024b}. It is also worth noting that the renaissance of selenium solar cells has only just begun: roughly 50 publications have reported complete devices with corresponding efficiencies. In this context, the generally observed efficiency vs effort trajectory for emerging PV absorbers \cite{dale2023a} is a highly encouraging indicator that we are only at the onset of the learning curve for selenium solar cells.\\

\section*{Acknowledgements}
RSN acknowledges Prof. Peter C. K. Vesborg, Prof. Emer. Ole Hansen, and Prof. Andrea Crovetto for years of useful discussions in the work on developing a new experimental platform for research on selenium as an optoelectronic material. The work presented here is supported by the Carlsberg Foundation, grant CF24-0200.


\nocite{*}

\bibliography{references}

\begin{figure*}[!t]
  \begin{minipage}[t]{0.30\textwidth}
    \vspace{0pt} 
    \includegraphics[width=\textwidth]{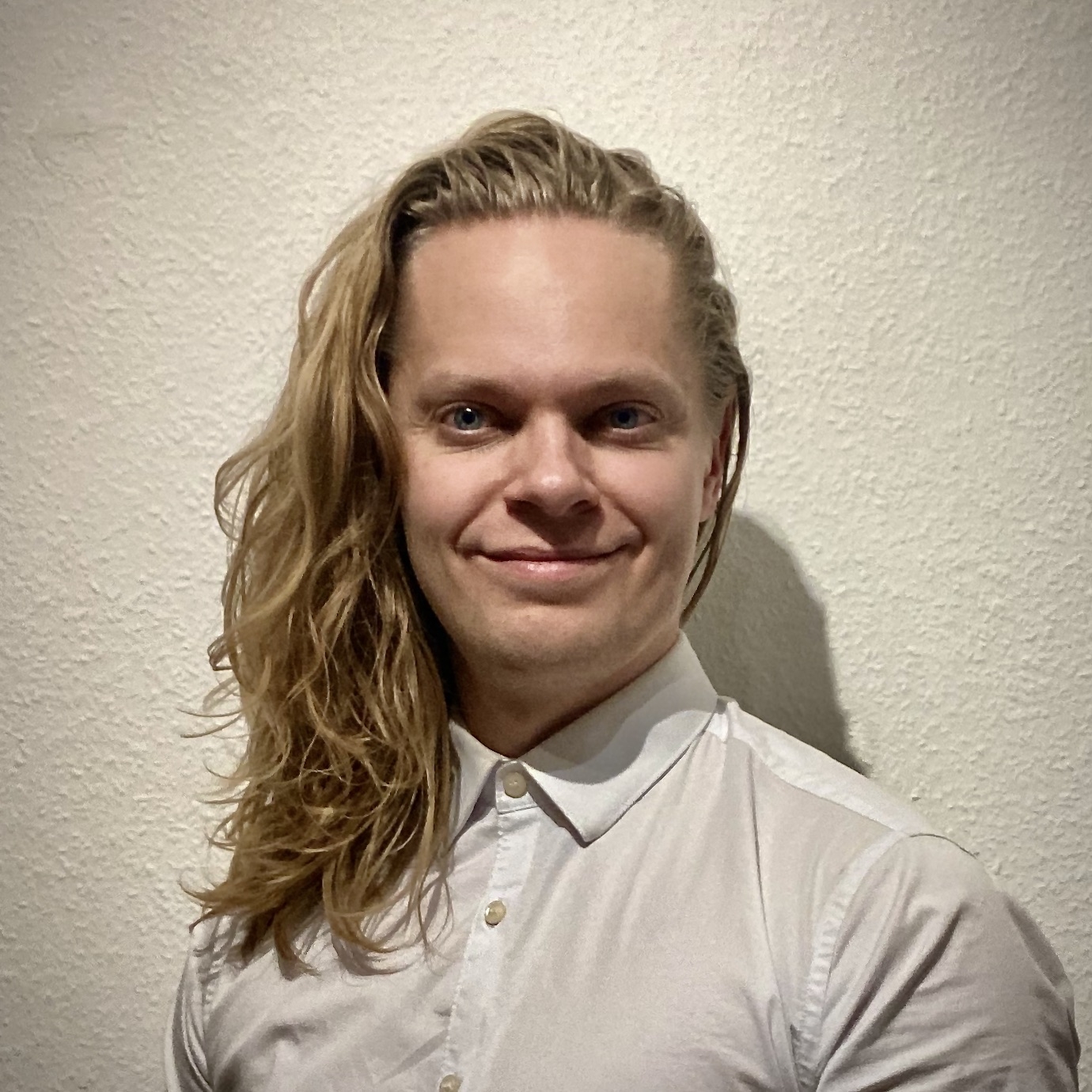}
  \end{minipage}%
  \hfill
  \begin{minipage}[t]{0.65\textwidth}
    \vspace{0pt} 
    \normalsize \justifying
    \textbf{Rasmus Svejstrup Nielsen} is a postdoctoral fellow at the Technical University of Denmark (DTU). He completed his Ph.D. in 2023 at DTU, focusing on optoelectronic semiconductor physics and novel materials for advanced photovoltaic concepts, including trigonal selenium as a wide bandgap photoabsorber for tandem solar cells. He then undertook a two-year international postdoctoral fellowship in the Nanomaterials Spectroscopy and Imaging Group at the Swiss Federal Laboratories for Materials Science and Technology (EMPA) in Dübendorf, Switzerland. Here, he explored the synthesis-structure-function relationships of inorganic chalcogenide perovskites and chalcohalides, which are considered promising for the next generation of solar cells. His main scientific interests are in energy materials, defect physics, and combinatorial/high-throughput materials discovery at the interface between computational and experimental science.
  \end{minipage}
\end{figure*}

\end{document}